# Saturated and Anisotropic Magnetostriction in an Altermagnet

*Zhiyuan Duan[1,8#], Qiyun Xu [2#], Peixin Qin[1,8]*, Li Liu[1,8], Guojian Zhao[1,8], Yuzhou He[3], Xiaoyang Tan[1,8], Sixu Jiang[1,8], Jingyu Li[1,8], Xiaoning Wang[4], Qinghua Zhang[3], Wenhui Duan[2,6,7], Yong Xu[2,6], Ziang Meng[1,8]*, Peizhe Tang[1,5]*, Chengbao Jiang[1,8]*, and Zhiqi Liu[1,8]**

[1]School of Materials Science and Engineering, Beihang University, Beijing, 100191, China.

[2]State Key Laboratory of Low Dimensional Quantum Physics and Department of Physics, Tsinghua University, Beijing, 100084, China.

[3]Beijing National Laboratory for Condensed Matter Physics, Institute of Physics, Chinese Academy of Sciences, Beijing, 100190, China.

[4]The Analysis & Testing Center, Beihang University, Beijing, 100191, China.

[5]Max Planck Institute for the Structure and Dynamics of Matter, Center for Free Electron Laser Science, Hamburg, 22761, Germany.

[6]Frontier Science Center for Quantum Information, Tsinghua University, Beijing, 100084, China.

[7]Institute for Advanced Study, Tsinghua University, Beijing, 100084, China.

[8]State Key Laboratory of Tropic Ocean Engineering Materials and Materials Evaluation, Beihang University, Beijing, 100191, China.

#These authors contributed equally to this work.

## ABSTRACT

Magnetostriction, a fundamental phenomenon bridging magnetism and mechanics, has enabled a broad spectrum of applications. For almost two centuries, it has been mainly investigated for ferromagnets. Regarding the magnetostriction of antiferromagnets (AFMs), limitedly known examples for both conventional collinear AFMs and noncollinear AFMs predominantly exhibit non-saturating magnetic-field dependence. Herein, we report an easily saturated magnetostriction effect in a prototypical altermagnet - MnTe, which is an emerging class of collinear AFMs with special crystal symmetries. For high-quality MnTe single crystals, the magnetostriction saturates under a moderate field of ~0.7 T with an intriguing two-fold-symmetry anisotropy. First-principles calculations reveal that the saturated and anisotropic magnetostriction originates from symmetry-allowed coupling between elastic strain and its Néel order parameter. These findings break the traditional wisdom on antiferromagnetic magnetostriction.

## INTRODUCTION

The coupling between magnetism and mechanics, manifesting as magnetostriction, serves as a cornerstone for various technologies ranging from sensors and actuators to precision micro-mechanical systems.[1–3] This phenomenon, historically rooted in ferromagnetic materials like Fe, Co, Ni and their alloys,[4,5] as well as giant magnetostrictive rare-earth compounds,[6–11] arises from the material's endeavor to minimize its total free energy under an applied magnetic field. In ferromagnets, the primary magnetostrictive response was attributed from magnetic domains,[2] which includes the magnetic field-induced displacement and rotation of magnetic domains, as well as the motion of domain walls. In contrast, the landscape of antiferromagnets (AFMs), despite their abundance over ferromagnets, remains largely uncharted. This disparity primarily roots from the inherently weaker magnetostrictive response in AFMs, which makes it experimentally more challenging to detect and study. Phenomenological observations on the few studied prototypes, such as the collinear AFM $MnF_2$,[12,13] CoO,[14] NiO [15,16] and the noncollinear AFM $Mn_3Sn$,[17–19] reveal not only a notably weaker magnetostrictive response but also a non-saturating trend under magnetic fields. The combination of a weak signal and the lack of a clear saturation threshold has limited the mechanistic understanding of antiferromagnetic magnetostriction, thereby spurring the search for exotic magnetic phases that, while possessing zero net magnetization, might exhibit stronger and more tractable magnetoelastic effects, potentially even a saturating magnetostriction akin to ferromagnets.

The recent emergence of altermagnets presents a transformative opportunity in this pursuit.[20–22] This new magnetic class uniquely combines vanishing net magnetization of AFMs with momentum-dependent spin-splitting characteristics reminiscent of ferromagnets, a feature stemming from specific crystal symmetries that lift spin degeneracy.[23,24] Theoretical predictions and subsequent experimental confirmations have identified a growing family of altermagnetic candidates.[25–31] These materials have already been demonstrated to host intriguing magnetoelectric and piezomagnetic phenomena,[32–40] establishing them as a promising platform for next-generation spintronics. However, their magnetomechanical properties—particularly magnetostriction—remain largely unexplored. This research gap is significant: unlike spin transport effects, magnetostriction directly reflects the spin-lattice interaction which is a fundamental coupling that plays a key role in determining the properties of a wide range of magnetic materials.[41–46] Therefore, the exploration of these phenomena would offer crucial insights into the role of antiferromagnetic order in altermagnets for the lattice response and supporting their application in magnetoelastic devices.

Building on the unique promise of altermagnets, MnTe stands out as a prototypical candidate, with its altermagnetic character firmly established by experimentally observed spin splitting band structure [23,24] and anomalous Hall effects.[34] Crystallizing in the hexagonal NiAs-type structure (space group $P6_3/mmc$, or $D_{6h}$), MnTe exhibits a Néel transition at ~307 K and displays metal-like electrical transport behavior across a broad temperature range (50–300 K).[24,47] Historically classified as a collinear AFM and *p*-type semiconductor,[24,48] the distinct altermagnetic nature of MnTe still calls for a re-examination of its magnetoelastic properties, such as the role of Néel vector for the magnetoelastic properties. Recent research has indeed predicted a coupling between Néel vector and strain tensor,[49] and several experimental studies have reported piezomagnetic effects in MnTe, demonstrating a stress-induced magnetization.[50–52] All these results suggest that MnTe exhibits robust magnetomechanical interactions. However, magnetostriction—the reverse effect of piezomagnetism—has not been systematically characterized in MnTe. This lack of data not only leaves the magnetoelastic response landscape of MnTe incomplete, but also hinders the validation of altermagnet-specific spin-lattice coupling mechanisms.

In this work, we report the direct observation of saturated and highly anisotropic magnetostriction in high-quality single-crystal MnTe. At cryogenic temperatures, the crystal exhibits a negative magnetostriction that saturates to about -30 ppm under a moderate field of ~0.7 T. Strikingly, the saturated magnetostriction ($\lambda_S$) within the (0001) plane reveals a pronounced two-fold symmetric dumbbell-shaped anisotropy, with its magnitude minimized along the $[01\bar{1}0]$ direction and maximized along the perpendicular $[2\bar{1}\bar{1}0]$ axis. Our theoretical calculations reveal that the symmetry-allowed coupling between elastic strain and the Néel order parameter in MnTe contributes to the saturated and anisotropic magnetostriction. The observed two-fold pattern for the anisotropic magnetostriction is governed by averaged contributions of differently oriented domains under in-plane magnetic field in a spin-flop process. This behavior not only provides direct evidence of magnetoelastic coupling in an altermagnet, but also inspires a theoretical understanding for magnetostriction in the altermagnetism with tunable Néel vector, underscoring a unique mechanism for spin-lattice interaction in zero-net-magnetization systems.

## RESULTS AND DISCUSSION

### High-quality MnTe single crystals

High-quality MnTe single crystals were synthesized using a modified Bridgman method. A schematic of growth setup is presented in Figure 1a, with the detailed procedure described in *Supporting Information*. The optical image of synthesized MnTe single crystals is shown in Figure 1b. The shiny surfaces were achieved by polishing with 2000-grit sandpaper. Powder X-ray diffraction (XRD) analysis (Figure 1c) confirms the phase purity of the sample, revealing only reflections of the $\alpha$-MnTe phase (space group $P6_3/mmc$ or $D_{6h}$). Rietveld refinement performed using GSAS-II yielded lattice parameters of $a$ = 4.1478 Å and $c$ = 6.7119 Å, consistent with previous crystallographic reports.[53,34,54]

In addition, atomic-resolution scanning transmission electron microscopy (STEM) was applied to characterize the MnTe single crystal. To aid the interpretation of the crystallographic directions, an atomic illustration showing the relationship between the Miller indices ([*uvw*]) and Miller-Bravais indices ([*uvtw*]) is provided in Figure S1. In Figure 2a–c, the hexagonal lattice symmetry of the synthesized crystals is directly revealed, indicating the high quality of our single crystal. As shown in Figure 2d, the electron energy loss spectroscopy (EELS) elemental mapping confirms a homogeneous and stoichiometric distribution of Mn and Te. Furthermore, high-angle annular dark-field (HAADF) imaging (Figure 2e–h) clearly resolves the individual Mn and Te atoms, exhibiting the characteristic (0001) crystal plane of the $\alpha$-MnTe phase. Since HAADF image contrast is approximately proportional to the square of the atomic number, the brighter spheres correspond to Te (higher atomic number), while the darker ones correspond to Mn. Collectively, these comprehensive structural characterizations confirm the successful synthesis of phase-pure, single-crystalline MnTe.

The samples employed for investigating the magnetic and electrical transport properties were obtained from as-grown single crystals. As shown in Figure S2, the electron backscatter diffraction (EBSD) inverse pole figure (IPF) maps display a uniform color for both Sample #1 and Sample #2, corresponding to the (0001) crystallographic orientation of the cleavage samples. No detectable grain boundaries are observed, indicating a single-crystalline nature. Moreover, the XRD patterns (Figure S3) show sharp (0002) and (0004) peak reflections, identifying the cleavage plane as the (0001) plane as well. The crystallographic axes were further determined by selected-area electron diffraction (SAED) in a transmission electron microscope, for which the specimens were prepared using focused ion beam (FIB) milling. In addition, the single-crystal XRD reveals the crystallographic directions. The in-plane crystallographic axes are clearly resolved, confirming the high crystallinity and well-defined surface orientation of the samples (Figure S4). Taken together, these orientation characterizations collectively confirm the precise crystallographic alignment of our MnTe single-crystal samples, laying a solid foundation for the subsequent measurements.

To characterize the altermagnetic to paramagnetic transition, we performed temperature-dependent resistivity and magnetization measurements. As shown in Figure 3a, the resistivity exhibits a distinct anomaly around ~307 K, in agreement with the previously reported altermagnetic-to-paramagnetic transition.[24,47,55] Beyond this transition, the resistivity continues to increase, albeit with a noticeable change in slope. As plotted in Figure 3b, magnetization measurements under a 0.1 T field applied along the $[2\bar{1}\bar{1}0]$ and [0001] axes reveal a consistent profile: both curves initially rise before decreasing above 307 K (marked by an arrow), signaling the Néel transition. Notably, magnetization along the [0001] axis is suppressed and smoother compared to that along $[2\bar{1}\bar{1}0]$, revealing pronounced magnetic anisotropy that indicates greater difficulty in magnetizing the crystal along the [0001] direction.

As further evidence of the altermagnetic character, the field-dependent magnetization was measured along the $[2\bar{1}\bar{1}0]$ and [0001] directions from 50 to 300 K under fields up to ±3 T. As summarized in Figure 3c, d, the field-dependent magnetic moment curves exhibit strictly linear behavior and pass through the origin at all temperatures, indicating a negligible net magnetization and the absence of macroscopic magnetic moment at zero field—both hallmarks of an antiferromagnetic ground state. The loops nearly overlap across the studied temperature range, showing only a minimal slope increase at elevated temperatures. A slight curvature is discernible in the low-field region of the $[2\bar{1}\bar{1}0]$ curve at 50 K, and that indicates the spin-flop transition in our sample. To amplify this trend, we plot the derivative of both $M$-$\mu_0 H$ curves. As shown in Figure 3e, f, the kink feature in the plot indicates that the spin-flop field of our MnTe single crystal is around 0.7 T, which not only accords with the previous study in MnTe epitaxy films,[56] but also closely matches the saturation field of the magnetostriction signal as measured afterwards. The derivative along the [0001] direction shows a similar trend, but the overall variation is smaller and smoother, indicating a linear behavior in the original $M$-$\mu_0 H$ curve, which suggests that the magnetic moments of MnTe lie within the (0001) plane.

**Saturated and anisotropic magnetostriction in altermagnetic MnTe**

We utilized specialized strain gauges capable of reliable operation at cryogenic temperatures for a more comprehensive characterization of MnTe's magnetostrictive behavior. The detailed experimental procedure is provided in *Experimental Section* (*Supporting Information*), with the complete set of raw data (of Sample #1) plotted in Figures S5-S12. For reproducibility, the pole figure of another specimen (Sample #2) is attached as Figure S13. Notably, magnetostriction measurements at a fixed angle across temperatures (50, 75, 100, 150, 200 and 250 K) reveal a clear saturation trend in all curves, with saturation achieved at an applied magnetic field of ~0.7 T (Figure 4), which highly matches the previous observation of spin-flop field.

In detail, as the external magnetic field increases, the MnTe crystal initially contracts along the field direction, exhibiting a negative magnetostriction. When the field reaches approximately 0.7 T, the magnetostriction gradually saturates to its maximum negative value and remains constant with further increase in the field. Upon reversing the field direction, the saturation behavior remains identical, demonstrating a symmetric dependence with respect to the magnetic field.

Whereas conventional collinear and noncollinear antiferromagnetic materials such as CoO,[14] $MnF_2$,[12,13] NiO,[16,57] and $Mn_3Sn$,[17–19] typically exhibit a nearly unsaturated negative magnetostriction within 1 T, the response in MnTe shows a distinct saturated profile. This observed nonlinearity suggests an underlying mechanism characteristic of altermagnets, as is well supported by the subsequent theoretical calculations.

This saturated response, however, becomes undetectable at room temperature. We attribute the absence of a discernible signal at 300 K to two possible factors: either the intrinsic magnetostriction of MnTe at room temperature falls below the resolution limit of our strain gauge, or the temperature-induced increase in the gauge's electrical resistance overwhelms the subtle resistance variation arising from the sample's magnetostriction.

When focusing on the in-plane magnetostriction within the (0001) plane, the $[2\bar{1}\bar{1}0]$ crystallographic direction was defined as angle $\theta = 0°$, with angles increasing clockwise. The magnetostriction was measured at 30° intervals from 0° to 180° at multiple temperatures, and the whole angular scan was repeated seven times. Figure 4a illustrates the measurement setup, and Figure 4b shows the magnetostriction curves measured at 50 K, allowing a direct comparison between angles. Since the magnetostriction response under negative magnetic fields is equivalent to that under positive fields at an angle shifted by 180°, and our preliminary measurements confirmed the reproducibility between positive and negative field directions, we used the data from angle $\theta$ to represent that from $\theta + 180°$ (*e.g.*, the 30° data also represents the 210° direction). Thus, the full 360° angular dependence is constructed by applying the structural equivalence between $\theta$ and $\theta + 180°$ to the experimental data measured between 0° and 180°, thereby reducing the systematic errors while maintaining physical consistency with the crystal symmetry. As shown in Figure 4c, the resulting angular-dependent pole figure of saturated magnetostriction ($\lambda_S$) exhibits a two-fold symmetric pattern resembling a dumbbell shape, demonstrating the anisotropic magnetostriction of MnTe in the (0001) plane. The maximum magnitude of $\lambda_S$ occurs at 0° (~33 ppm at 50 K for $[2\bar{1}\bar{1}0]$ axis), while the minimum magnitude emerges along the perpendicular direction (~20 ppm at 50 K for $[01\bar{1}0]$ axis). The characteristic dumbbell-shaped anisotropy remains clearly observable up to 200 K. As the temperature increases to 250 K, the amplitude of the magnetostriction diminishes, and the signal-to-noise ratio is compromised by the limited resolution of the strain gauge. While the angular dependence becomes ambiguous under these conditions, the data are consistent with a gradual weakening of the effect and do not indicate a phase transition.

**Theoretical framework and mechanism**

To study the underlying physics for the observed phenomena, we employ a symmetry-based Landau theory to understand the saturated, anisotropic magnetostriction in MnTe. The total free energy of this system is written as $F = F_{lattice} + F_{mec}$. When we focus on the (0001) in-plane magnetostriction, the elastic energy $F_{lattice}$ of an in-plane strained hexagonal crystal is given by:[44]

$$F_{lattice} = \frac{c_{11}}{4}\left(\epsilon_B^2 + \epsilon_\mu^2\right) + \frac{c_{12}}{4}\left(\epsilon_B^2 - \epsilon_\mu^2\right) + (c_{11} - c_{12})\epsilon_{xy}^2 \quad (1)$$

where $\epsilon_{ij}$ are the components of the strain tensor and $c_{ij}$ is the elastic stiffness tensor. Herein, $\epsilon_B \equiv \epsilon_{xx} + \epsilon_{yy}$ is the in-plane isotropic strain and $\epsilon_\mu \equiv \epsilon_{xx} - \epsilon_{yy}$ is the in-plane anisotropic strain. As discussed in *Experimental Section* (*Supporting Information*), the coupling terms $F_{mec}$ allowed by the point-group symmetry are:

$$F_{mec} = -\lambda_1\epsilon_B\left(n_x^2 + n_y^2\right) - \lambda_2\left[\epsilon_\mu\left(n_x^2 - n_y^2\right) + 4\epsilon_{xy}n_xn_y\right] \quad (2)$$

where $n_i$ are the components of the staggered magnetization as the Néel order parameter, and $\lambda_i$ is the coupling constant. Note that $\epsilon_{ij}$ determines the change of the length $(\Delta L/L)_{\hat{l}} = \sum_{ij}\epsilon_{ij} l_i l_j$ along the detection direction that has an angle of $\theta$ with $[2\bar{1}\bar{1}0]$ axis, where unit vector $\hat{l} = cos\theta\hat{x} + sin\theta\hat{y}$. Minimizing total free energy $F$ with respect to the change of applied strain, we find

$$(\Delta L/L)_{\hat{l}} = \frac{\epsilon_B}{2} + \frac{\epsilon_\mu}{2}\cos2\theta + \epsilon_{xy}\sin2\theta = N_0^2\left[\frac{\lambda_1}{c_{11}+c_{12}} + \frac{\lambda_2}{c_{11}-c_{12}}\cos 2(\theta-\psi)\right] \quad (3)$$

where $N_0$ denotes the magnitude and $\psi$ is the orientation angle of the Néel vector in a single domain. The coupling constant and the stiffness tensor are obtained from density-functional theory (DFT) calculations (see details in *Experimental Section* (*Supporting Information*)). By applying an in-plane magnetic field, the field-induced additional energy term could compete with the intrinsic magnetocrystalline anisotropy, as reported previously.[58] Thus, the potential magnetic energy surface can be effectively tuned. As the magnetic field is increased from zero, the Néel vector rotates away from its initial orientation. Once the magnetic field is large enough to induce the spin-flop transition, ultimately the Néel vector culminates to the perpendicular direction of the applied magnetic field (the critical spin-flop field $h_{sf}$). During the process, the angle in Eq. (3) evolves from $\psi = \psi_0$ to $\psi = \theta + \pi/2$, with saturation achieved. The corresponding length change is

$$\Delta((\Delta L/L)_{\hat{l}}) = (\Delta L/L)_{\hat{l}}|_{h=h_{sf}} - (\Delta L/L)_{\hat{l}}|_{h=0} = -\frac{\lambda_2 N_0^2}{c_{11}-c_{12}}[1+\cos2(\theta-\psi_0)] \quad (4)$$

We have analyzed the single-domain case discussed above and present the pole figure in Figure S14.

It is worth noting that, previous studies on MnTe thin films and bulk crystals collectively demonstrate the characteristics of a multidomain state,[59–61] which is further confirmed by magnetic force microscopy (MFM) in our work (Figure S15). As shown in Figure 5a, due to the $D_{6\mathrm{h}}$ symmetry of MnTe, six symmetry-equivalent domains coexist with different magnetic order parameters marked by different colors indicating distinct dipole orientations. At zero magnetic field, the total domain-averaged Néel vector remains finite, indicating that the six orientations are not equally populated. Recent circular-dichroism measurements in MnTe substantiate the multidomain picture and quantify the relative domain populations.[62,63] In our simplified model, domain-wall contributions are neglected, and specimen-size effects are not explicitly parameterized. We apply an external magnetic field along the direction which has an angle of $\theta$ with $[2\bar{1}\bar{1}0]$ axis. Correspondingly, the applied magnetic field could progressively cant the Néel vector in each domain.

As illustrated in Figure 5b, the magnetostriction in each domain reaches saturation at a critical field, at which the Néel vectors in all domains become aligned perpendicular to the direction of the applied magnetic field. The total magnetostrictive response of the sample is the area-weighted sum of the contributions from all individual domains. As shown in Figure 5c, the saturated magnetostriction is anisotropic with the dumbbell shape for the change of $\theta$, which is robust with respect to variations in the domain fractions. In MnTe, the spin-flop-related field-driven spin reorientation dominates the magnetostriction response within the measured field window, leading to the observed saturation trend. Further analysis of the robustness against domain fractions variation is provided in Figure S16. The length change amplitude $A = \lambda_2 N_0^2/(c_{11} - c_{12})$ obtained from the DFT calculation is around 28.8 ppm (see details in *Experimental Section* (*Supporting Information*)), which is on the same order of magnitude as the value measured experimentally at low temperature. Our theoretical analysis and calculations provide insight into the microscopic mechanisms underlying magnetostriction in the altermagnet MnTe, establishing a general scenario for investigating magnetoelastic responses in antiferromagnetic materials.

## CONCLUSION

In summary, the magnetostrictive behavior of altermagnetic MnTe exhibits unique characteristics. At cryogenic temperatures, the single-crystal MnTe displays a saturated negative magnetostriction under external magnetic fields, contrasting with the unsaturated response of conventional AFMs. A striking two-fold symmetric anisotropy of the saturated magnetostriction $\lambda_S$ was observed across different crystallographic directions within the (0001) plane, with its magnitude maximized along the $[2\bar{1}\bar{1}0]$ direction and minimized along the $[01\bar{1}0]$ direction. The Landau theory analysis and DFT calculations confirm that these phenomena arise from coupling between elastic strain and the Néel vector components and, at the spin-flop saturation field, reflect the collective response of multiple domains. This study enriches the understanding of magnetoelastic effects in altermagnetic systems. Importantly, magnetostriction in such systems avoids the hysteresis and energy loss inherent to traditional magnetostrictive materials, enabling potential ultrafast response and high-density integration. Our work positions altermagnets as promising candidates for zero-net-magnetization magnetostrictive device applications.

## ASSOCIATED CONTENT

**Supporting Information**.

The Supporting Information is available free of charge at Detailed experimental procedures, crystal structure illustrations, EBSD maps, XRD patterns, SC-XRD images, MFM topographic, magnetostriction results at various temperatures and angles, pole figures, multidomain simulations, and DFT-calculated total free energies (PDF)

**Data Availability Statement**

All the data and simulation codes in this study are available from the corresponding authors upon request.

**AUTHOR INFORMATION**

**Corresponding Author**

*Emails: qinpeixin@buaa.edu.cn; mengza@buaa.edu.cn; peizhet@buaa.edu.cn; jiangcb@buaa.edu.cn; zhiqi@buaa.edu.cn

**Author Contributions**

Z.D. and Q.X. contributed equally to this work. The manuscript was written through contributions of all authors. All authors have given approval to the final version of the manuscript.

**Notes**

The authors declare no competing financial interest.

**ACKNOWLEDGMENTS**

Z.L. acknowledges financial support from the National Key R&D Program of China (no. 2022YFA1602700). Z.L. acknowledges financial support from the Beijing Natural Science Foundation (no. JQ23005). P.Q. acknowledges financial support from the National Natural Science Foundation of China (no. 52401300). Z.M. acknowledges financial support from the National Natural Science Foundation of China (no. 524B2003). L.L. acknowledges financial support from the National Natural Science Foundation of China (no. 525B2008). Z.L. acknowledges financial support from the National Natural Science Foundation of China (nos. 52425106and 52271235). Z.L. and C.J. acknowledge financial support from the National Natural Science Foundation of China (no. 52121001). This work is supported by National Natural Science Foundation of China (no. U25A20244). P.T. acknowledges financial support from the National Natural Science Foundation of China (nos. 12234011 and 12374053). P.T. and Y.X. acknowledge financial support from the National Key R&D Program of China (no. 2024YFA1409100). Y.X. acknowledges financial support from the National Key R&D Program of China (no. 2023YFA1406400). Y.X. acknowledges financial support from the National Natural Science Foundation of China (nos. 12334003, 12421004 and 12361141826). Y.X. acknowledges financial support from the National Science Fund for Distinguished Young Scholars (no. 12025405). W.D. acknowledges financial support from the Innovation Program for Quantum Science and Technology (no. 2023ZD0300500). Y.X. and W.D. acknowledge financial support from the Basic Science Center Project of NSFC (no. 52388201). Q.Z. acknowledges funding from the National Natural Science Foundation of China (no. 52322212) and from the National Key R&D Program of China (nos. 2023YFA1406300 and 2024YFA1409500). This work is supported by the Fundamental Research Funds for the Central Universities. The authors acknowledge the Analysis & Testing Center of Beihang University for the assistance.

## REFERENCES

(1) Joule, J. P. XVII. On the Effects of Magnetism upon the Dimensions of Iron and Steel Bars. *Lond. Edinb. Phil. Mag.* **1847**, *30* (199), 76–87.

(2) Chandrasekhar, B. s.; Fawcett, E. Magnetostriction in Metals. *Adv. Phys.* **1971**, *20* (88), 775–794.

(3) Engdahl, G.; Bright, C. B. Chapter 5 - Device Application Examples. In *Handbook of Giant Magnetostrictive Materials*; Engdahl, G., Ed.; Electromagnetism; Academic Press: San Diego, 2000; pp 287–322.

(4) McCorkle, P. Magnetostriction and Magnetoelectric Effects in Iron, Nickel and Cobalt. *Phys. Rev.* **1923**, *22* (3), 271–278.

(5) Xing, Q.; Lograsso, T. A.; Ruffoni, M. P.; Azimonte, C.; Pascarelli, S.; Miller, D. J. Experimental Exploration of the Origin of Magnetostriction in Single Crystalline Iron. *Appl. Phys. Lett.* **2010**, *97* (7), 072508.

(6) Verhoeven, J. D.; Ostenson, J. E.; Gibson, E. D.; McMasters, O. D. The Effect of Composition and Magnetic Heat Treatment on the Magnetostriction of $Tb_xDy_{1-x}Fe_y$ Twinned Single Crystals. *J. Appl. Phys.* **1989**, *66* (2), 772–779.

(7) Wang, N.; Liu, Y.; Zhang, H.; Chen, X.; Li, Y. Fabrication, Magnetostriction Properties and Applications of Tb-Dy-Fe Alloys: A Review. *China Foundry* **2016**, *13* (2), 75–84.

(8) Cullen, J. R.; Clark, A. E.; Wun-Fogle, M.; Restorff, J. B.; Lograsso, T. A. Magnetoelasticity of Fe–Ga and Fe–Al Alloys. *J. Magn. Magn. Mater.* **2001**, *226–230*, 948–949.

(9) Kellogg, R. A.; Russell, A. M.; Lograsso, T. A.; Flatau, A. B.; Clark, A. E.; Wun-Fogle, M. Tensile Properties of Magnetostrictive Iron–Gallium Alloys. *Acta Mater.* **2004**, *52* (17), 5043–5050.

(10) Xing, Q.; Du, Y.; McQueeney, R. J.; Lograsso, T. A. Structural Investigations of Fe–Ga Alloys: Phase Relations and Magnetostrictive Behavior. *Acta Mater.* **2008**, *56* (16), 4536–4546.

(11) He, Y.; Ke, X.; Jiang, C.; Miao, N.; Wang, H.; Coey, J. M. D.; Wang, Y.; Xu, H. Interaction of Trace Rare-Earth Dopants and Nanoheterogeneities Induces Giant Magnetostriction in Fe-Ga Alloys. *Adv. Funct. Mater.* **2018**, *28* (20), 1800858.

(12) Matolyak, J.; Seehra, M. S.; Pavlovic, A. S. Magnetostriction in $MnF_2$. *Phys. Lett. A* **1974**, *49* (4), 333–334.

(13) Semenenko, L. M.; Dudko, K. L. Magnetostriction of Antiferromagnetic $MnF_2$ in a Strong Field. *Sov. J. Low Temp. Phys.* **1975**, *1* (9), 541–544.

(14) Nakamichi, T. Magnetostrictive Behavior of Antiferromagnetic CoO Single Crystal in Magnetic Field. *J. Phys. Soc. Jpn.* **1965**, *20* (5), 720–726.

(15) Yamada, T. Magnetic Anisotropy, Magnetostriction, and Magnetic Domain Walls in NiO. I. Theory. *J. Phys. Soc. Jpn.* **1966**, *21* (4), 664–671.

(16) Weber, N. B. Magnetostrictive Domain Walls in Antiferromagnetic NiO. *Phys. Rev. Lett.* **2003**, *91* (23), 237205.

(17) Pradenas, B.; Tchernyshyov, O. Spin-Frame Field Theory of a Three-Sublattice Antiferromagnet. *Phys. Rev. Lett.* **2024**, *132* (9), 096703.

(18) Meng, Q.; Dong, J.; Nie, P.; Xu, L.; Wang, J.; Jiang, S.; Zuo, H.; Zhang, J.; Li, X.; Zhu, Z.; Balents, L.; Behnia, K. Magnetostriction, Piezomagnetism and Domain Nucleation in a Kagome Antiferromagnet. *Nat. Commun.* **2024**, *15* (1), 6921.

(19) Wang, X.; Dong, W.; Qin, P.; Liu, J.; Jiang, D.; Zhao, S.; Duan, Z.; Zhou, X.; Chen, H.; Meng, Z.; Liu, L.; Zhao, G.; Xia, Z.; Zuo, H.; Zhu, Z.; Wang, J.; Xu, Y.; Kang, D.; Zhang, Q.; Tang, P.; Jiang, C.; Liu, Z. Giant Non-Saturating Exchange Striction in a Noncollinear Antiferromagnet. *Adv. Mater.* **2025**, *37* (18), 2500829.

(20) Šmejkal, L.; González-Hernández, R.; Jungwirth, T.; Sinova, J. Crystal Time-Reversal Symmetry Breaking and Spontaneous Hall Effect in Collinear Antiferromagnets. *Sci. Adv.* **2020**, *6* (23), eaaz8809.

(21) Šmejkal, L.; Sinova, J.; Jungwirth, T. Emerging Research Landscape of Altermagnetism. *Phys. Rev. X* **2022**, *12* (4), 040501.

(22) Šmejkal, L.; Sinova, J.; Jungwirth, T. Beyond Conventional Ferromagnetism and Antiferromagnetism: A Phase with Nonrelativistic Spin and Crystal Rotation Symmetry. *Phys. Rev. X* **2022**, *12* (3), 031042.

(23) Lee, S.; Lee, S.; Jung, S.; Jung, J.; Kim, D.; Lee, Y.; Seok, B.; Kim, J.; Park, B. G.; Šmejkal, L.; Kang, C.-J.; Kim, C. Broken Kramers Degeneracy in Altermagnetic MnTe. *Phys. Rev. Lett.* **2024**, *132* (3), 036702.

(24) Krempaský, J.; Šmejkal, L.; D'Souza, S. W.; Hajlaoui, M.; Springholz, G.; Uhlířová, K.; Alarab, F.; Constantinou, P. C.; Strocov, V.; Usanov, D.; Pudelko, W. R.; González-Hernández, R.; Birk Hellenes, A.; Jansa, Z.; Reichlová, H.; Šobáň, Z.; Gonzalez Betancourt, R. D.; Wadley, P.; Sinova, J.; Kriegner, D.; Minár, J.; Dil, J. H.; Jungwirth, T. Altermagnetic Lifting of Kramers Spin Degeneracy. *Nature* **2024**, *626* (7999), 517–522.

(25) Feng, Z.; Zhou, X.; Šmejkal, L.; Wu, L.; Zhu, Z.; Guo, H.; González-Hernández, R.; Wang, X.; Yan, H.; Qin, P.; Zhang, X.; Wu, H.; Chen, H.; Meng, Z.; Liu, L.; Xia, Z.; Sinova, J.; Jungwirth, T.; Liu, Z. An Anomalous Hall Effect in Altermagnetic Ruthenium Dioxide. *Nat. Electron.* **2022**, *5* (11), 735–743.

(26) Reimers, S.; Odenbreit, L.; Šmejkal, L.; Strocov, V. N.; Constantinou, P.; Hellenes, A. B.; Jaeschke Ubiergo, R.; Campos, W. H.; Bharadwaj, V. K.; Chakraborty, A.; Denneulin, T.; Shi, W.; Dunin-Borkowski, R. E.; Das, S.; Kläui, M.; Sinova, J.; Jourdan, M. Direct Observation of Altermagnetic Band Splitting in CrSb Thin Films. *Nat. Commun.* **2024**, *15* (1), 2116.

(27) Reichlova, H.; Lopes Seeger, R.; González-Hernández, R.; Kounta, I.; Schlitz, R.; Kriegner, D.; Ritzinger, P.; Lammel, M.; Leiviskä, M.; Birk Hellenes, A.; Olejník, K.; Petřiček, V.; Doležal, P.; Horak, L.; Schmoranzerova, E.; Badura, A.; Bertaina, S.; Thomas, A.; Baltz, V.; Michez, L.; Sinova, J.; Goennenwein, S. T. B.; Jungwirth, T.; Šmejkal, L. Observation of a Spontaneous Anomalous Hall Response in the $Mn_5Si_3$ *d*-Wave Altermagnet Candidate. *Nat. Commun.* **2024**, *15* (1), 4961.

(28) Jiang, B.; Hu, M.; Bai, J.; Song, Z.; Mu, C.; Qu, G.; Li, W.; Zhu, W.; Pi, H.; Wei, Z.; Sun, Y.-J.; Huang, Y.; Zheng, X.; Peng, Y.; He, L.; Li, S.; Luo, J.; Li, Z.; Chen, G.; Li, H.; Weng, H.; Qian, T. A Metallic Room-Temperature d-Wave Altermagnet. *Nat. Phys.* **2025**, *21* (5), 754–759.

(29) Zhang, F.; Cheng, X.; Yin, Z.; Liu, C.; Deng, L.; Qiao, Y.; Shi, Z.; Zhang, S.; Lin, J.; Liu, Z.; Ye, M.; Huang, Y.; Meng, X.; Zhang, C.; Okuda, T.; Shimada, K.; Cui, S.; Zhao, Y.; Cao, G.-H.; Qiao, S.; Liu, J.; Chen, C. Crystal-Symmetry-Paired Spin–Valley Locking in a Layered Room-Temperature Metallic Altermagnet Candidate. *Nat. Phys.* **2025**, *21* (5), 760–767.

(30) Hu, X.; Zhao, W.; Xia, W.; Sun, H.; Wu, C.; Wu, Y.-Z.; Li, P. Valley Polarization and Anomalous Valley Hall Effect in Altermagnet $Ti_2Se_2S$ with Multipiezo Properties. *Appl. Phys. Lett.* **2025**, *127* (1), 011905.

(31) Oishi, R.; Taniguchi, T.; Adroja, D. T.; Le, M. D.; Aouane, M.; Onimaru, T.; Umeo, K.; Ishii, I.; Takabatake, T. $TbPt_6Al_3$: A Rare-Earth Based *g*-Wave Altermagnet with a Honeycomb Structure. *Phys. Rev. B* **2025**, *112* (9), 094421.

(32) Tschirner, T.; Keßler, P.; Gonzalez Betancourt, R. D.; Kotte, T.; Kriegner, D.; Büchner, B.; Dufouleur, J.; Kamp, M.; Jovic, V.; Smejkal, L.; Sinova, J.; Claessen, R.; Jungwirth, T.; Moser, S.; Reichlova, H.; Veyrat, L. Saturation of the Anomalous Hall Effect at High Magnetic Fields in Altermagnetic $RuO_2$. *APL Mater.* **2023**, *11* (10), 101103.

(33) Wang, M.; Tanaka, K.; Sakai, S.; Wang, Z.; Deng, K.; Lyu, Y.; Li, C.; Tian, D.; Shen, S.; Ogawa, N.; Kanazawa, N.; Yu, P.; Arita, R.; Kagawa, F. Emergent Zero-Field Anomalous Hall Effect in a Reconstructed Rutile Antiferromagnetic Metal. *Nat. Commun.* **2023**, *14* (1), 8240.

(34) Gonzalez Betancourt, R. D.; Zubáč, J.; Gonzalez-Hernandez, R.; Geishendorf, K.; Šobáň, Z.; Springholz, G.; Olejník, K.; Šmejkal, L.; Sinova, J.; Jungwirth, T.; Goennenwein, S. T. B.; Thomas, A.; Reichlová, H.; Železný, J.; Kriegner, D. Spontaneous Anomalous Hall Effect Arising from an Unconventional Compensated Magnetic Phase in a Semiconductor. *Phys. Rev. Lett.* **2023**, *130* (3), 036702.

(35) Duan, Z.; Qin, P.; Zhong, C.; Zhang, S.; Liu, L.; Zhao, G.; Wang, X.; Chen, H.; Meng, Z.; Li, J.; Jiang, S.; Tan, X.; Wu, Q.; Liu, Y.; Liu, Z. Electric-Field-Controlled Altermagnetic Transition for Neuromorphic Computing. *J. Am. Chem. Soc.* **2025**, *147* (51), 47330–47338.

(36) Bose, A.; Schreiber, N. J.; Jain, R.; Shao, D.-F.; Nair, H. P.; Sun, J.; Zhang, X. S.; Muller, D. A.; Tsymbal, E. Y.; Schlom, D. G.; Ralph, D. C. Tilted Spin Current Generated by the Collinear Antiferromagnet Ruthenium Dioxide. *Nat. Electron.* **2022**, *5* (5), 267–274.

(37) Bai, H.; Han, L.; Feng, X. Y.; Zhou, Y. J.; Su, R. X.; Wang, Q.; Liao, L. Y.; Zhu, W. X.; Chen, X. Z.; Pan, F.; Fan, X. L.; Song, C. Observation of Spin Splitting Torque in a Collinear Antiferromagnet $RuO_2$. *Phys. Rev. Lett.* **2022**, *128* (19), 197202.

(38) Karube, S.; Tanaka, T.; Sugawara, D.; Kadoguchi, N.; Kohda, M.; Nitta, J. Observation of Spin-Splitter Torque in Collinear Antiferromagnetic $RuO_2$. *Phys. Rev. Lett.* **2022**, *129* (13), 137201.

(39) Noh, S.; Kim, G.-H.; Lee, J.; Jung, H.; Seo, U.; So, G.; Lee, J.; Lee, S.; Park, M.; Yang, S.; Oh, Y. S.; Jin, H.; Sohn, C.; Yoo, J.-W. Tunneling Magnetoresistance in Altermagnetic $RuO_2$-Based Magnetic Tunnel Junctions. *Phys. Rev. Lett.* **2025**, *134* (24), 246703.

(40) Chen, H.; Wang, Z.-A.; Qin, P.; Meng, Z.; Zhou, X.; Wang, X.; Liu, L.; Zhao, G.; Duan, Z.; Zhang, T.; Liu, J.; Shao, D.-F.; Jiang, C.; Liu, Z. Spin-Splitting Magnetoresistance in Altermagnetic $RuO_2$ Thin Films. *Adv. Mater.* **2025**, 2507764.

(41) Liu, J.; Zhou, L.; Liang, R.; Li, S.; Li, Z.; Zhang, L.; Li, Z.; Wang, H.; Huang, Y.; Liu, R.; Tang, N. Tunable Carr-Type Temperature Dependence of Uniaxial Magnetocrystalline Anisotropy in Fe-Deficient $Fe_{3-x}GeTe_2$. *Phys. Rev. B* **2025**, *111* (14), 144425.

(42) Liu, J.; Jiang, J.; Wang, L.; Zhou, L.; Liang, R.; Li, S.; Li, Z.; Chang, K.; Yang, H.; Tang, N. Exotic Temperature Dependence of Uniaxial Magnetocrystalline Anisotropy in a Two-Dimensional Ferromagnet. *Phys. Rev. Lett.* **2025**, *134* (11), 116702.

(43) Liu, J.; Zhou, L.; Li, S.; Zhao, E.; Tang, N. Anomalous Temperature Dependence of Uniaxial Magnetocrystalline Anisotropy in the van Der Waals Ferromagnet $Fe_3GeTe_2$. *Phys. Rev. B* **2024**, *109* (6), L060404.

(44) Liang, R.; Zhou, L.; Liu, J.; Tang, N. Anisotropic Magnetic Critical Behavior of van Der Waals Room-Temperature Ferromagnet $Fe_5GeTe_2$ Identified by Magnetocaloric Measurements. *J. Appl. Phys.* **2024**, *136* (19), 193902.

(45) Liang, R.; Liu, J.; Zhou, L.; Tang, N. Large Relative Cooling Power in van Der Waals Room-Temperature Ferromagnet $Fe_5GeTe_2$. *Appl. Phys. Lett.* **2024**, *124* (18), 182406.

(46) Li, Z.; Li, S.; Xu, Y.; Tang, N. Recent Advances in Magnetism of Graphene from 0D to 2D. *Chem. Commun.* **2023**, *59* (42), 6286–6300.

(47) Kriegner, D.; Výborný, K.; Olejník, K.; Reichlová, H.; Novák, V.; Marti, X.; Gazquez, J.; Saidl, V.; Němec, P.; Volobuev, V. V.; Springholz, G.; Holý, V.; Jungwirth, T. Multiple-Stable Anisotropic Magnetoresistance Memory in Antiferromagnetic MnTe. *Nat. Commun.* **2016**, *7* (1), 11623.

(48) Ferrer-Roca, Ch.; Segura, A.; Reig, C.; Muñoz, V. Temperature and Pressure Dependence of the Optical Absorption in Hexagonal MnTe. *Phys. Rev. B* **2000**, *61* (20), 13679–13686.

(49) McClarty, P. A.; Rau, J. G. Landau Theory of Altermagnetism. *Phys. Rev. Lett.* **2024**, *132* (17), 176702.

(50) Baral, R.; Abeykoon, A. M. M.; Campbell, B. J.; Frandsen, B. A. Giant Spontaneous Magnetostriction in MnTe Driven by a Novel Magnetostructural Coupling Mechanism. *Adv. Funct. Mater.* **2023**, *33* (46), 2305247.

(51) Devaraj, N.; Bose, A.; Narayan, A. Interplay of Altermagnetism and Pressure in Hexagonal and Orthorhombic MnTe. *Phys. Rev. Mater.* **2024**, *8* (10), 104407.

(52) Yershov, K. V.; Gomonay, O.; Sinova, J.; Van Den Brink, J.; Kravchuk, V. P. Curvature-Induced Magnetization of Altermagnetic Films. *Phys. Rev. Lett.* **2025**, *134* (11), 116701.

(53) De Melo, O.; Leccabue, F.; Pelosi, C.; Sagredo, V.; Chourio, M.; Martin, J.; Bocelli, G.; Calestani, G. Crystal Growth and Characterization of MnTe Single Crystals. *J. Cryst. Growth* **1991**, *110* (3), 445–451.

(54) Belashchenko, K. D. Giant Strain-Induced Spin Splitting Effect in MnTe, a *g*-Wave Altermagnetic Semiconductor. *Phys. Rev. Lett.* **2025**, *134* (8), 086701.

(55) Aoyama, T.; Ohgushi, K. Piezomagnetic Properties in Altermagnetic MnTe. *Phys. Rev. Mater.* **2024**, *8* (4), L041402.

(56) Kriegner, D.; Reichlova, H.; Grenzer, J.; Schmidt, W.; Ressouche, E.; Godinho, J.; Wagner, T.; Martin, S. Y.; Shick, A. B.; Volobuev, V. V.; Springholz, G.; Holý, V.; Wunderlich, J.; Jungwirth, T.; Výborný, K. Magnetic Anisotropy in Antiferromagnetic Hexagonal MnTe. *Phys. Rev. B* **2017**, *96* (21), 214418.

(57) Kresse, G.; Furthmüller, J. Efficient Iterative Schemes for Ab Initio Total-Energy Calculations Using a Plane-Wave Basis Set. *Phys. Rev. B* **1996**, *54* (16), 11169–11186.

(58) Landau, L. D.; Pitaevskii, L. P.; Kosevich, A. M.; Lifshitz, E. M. *Theory of Elasticity: Volume 7*; Elsevier, 2012; Vol. 7.

(59) Amin, O. J.; Dal Din, A.; Golias, E.; Niu, Y.; Zakharov, A.; Fromage, S. C.; Fields, C. J. B.; Heywood, S. L.; Cousins, R. B.; Maccherozzi, F.; Krempasky, J.; Dil, J. H.; Kriegner,

D.; Kiraly, B.; Campion, R. P.; Rushforth, A. W.; Edmonds, K. W.; Dhesi, S. S.; Smejkal, L.; Jungwirth, T.; Wadley, P. Nanoscale Imaging and Control of Altermagnetism in MnTe. *Nature* **2024**, *636* (8042), 348–353.

(60) Hariki, A.; Dal Din, A.; Amin, O. J.; Yamaguchi, T.; Badura, A.; Kriegner, D.; Edmonds, K. W.; Campion, R. P.; Wadley, P.; Backes, D.; Veiga, L. S. I.; Dhesi, S. S.; Springholz, G.; Smejkal, L.; Vyborny, K.; Jungwirth, T.; Kunes, J. X-Ray Magnetic Circular Dichroism in Altermagnetic $\alpha$-MnTe. *Phys. Rev. Lett.* **2024**, *132* (17), 176701.

(61) Yamamoto, R.; Turnbull, L.; Di Pietro Martínez, M.; Ferreira Raboni, M.; Štefančič, A.; Mayoh, D. A.; Balakrishnan, G.; Pei, Z.; Xue, P.; Chang, L.; Ringe, E.; Harrison, R.; Valencia, S.; Kazemian, M.; Kaulich, B.; Donnelly, C. Altermagnetic Nanotextures Revealed in Bulk MnTe. *Phys. Rev. Appl.* **2025**, *24* (3), 034037.

(62) Takegami, D.; Aoyama, T.; Okauchi, T.; Yamaguchi, T.; Tippireddy, S.; Agrestini, S.; García-Fernández, M.; Mizokawa, T.; Ohgushi, K.; Zhou, K.-J.; Chaloupka, J.; Kuneš, J.; Hariki, A.; Suzuki, H. Circular Dichroism in Resonant Inelastic X-Ray Scattering: Probing Altermagnetic Domains in MnTe. *Phys. Rev. Lett.* **2025**, *135* (19), 196502.

(63) Krüger, P. Circular Dichroism in Resonant Photoelectron Diffraction as a Direct Probe of Sublattice Magnetization in Altermagnets. *Phys. Rev. Lett.* **2025**, *135* (19), 196703.

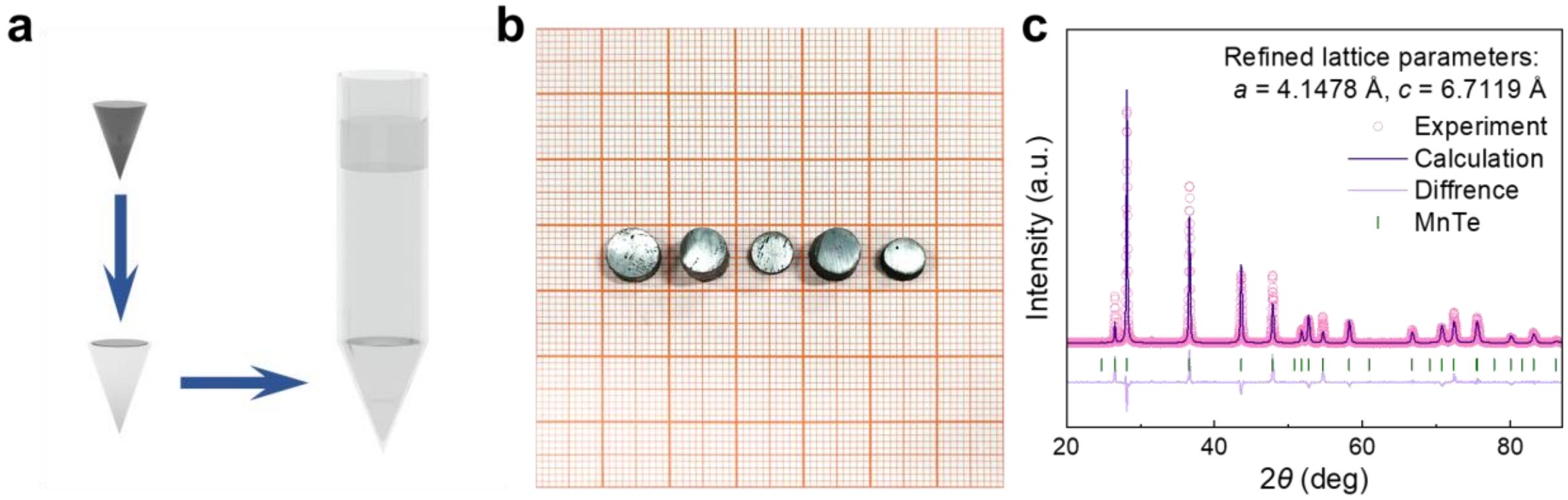

**Figure 1.** Synthesis, Morphology, and Structural Analysis of MnTe Single Crystals. (a) Illustration of the assembly process of quartz tube. (b) Optical image of the cut and polished MnTe single crystals. (c) XRD curve of ground MnTe with Rietveld refinement results with lattice parameters: $a$ = 4.1478 Å, $c$ = 6.7119 Å).

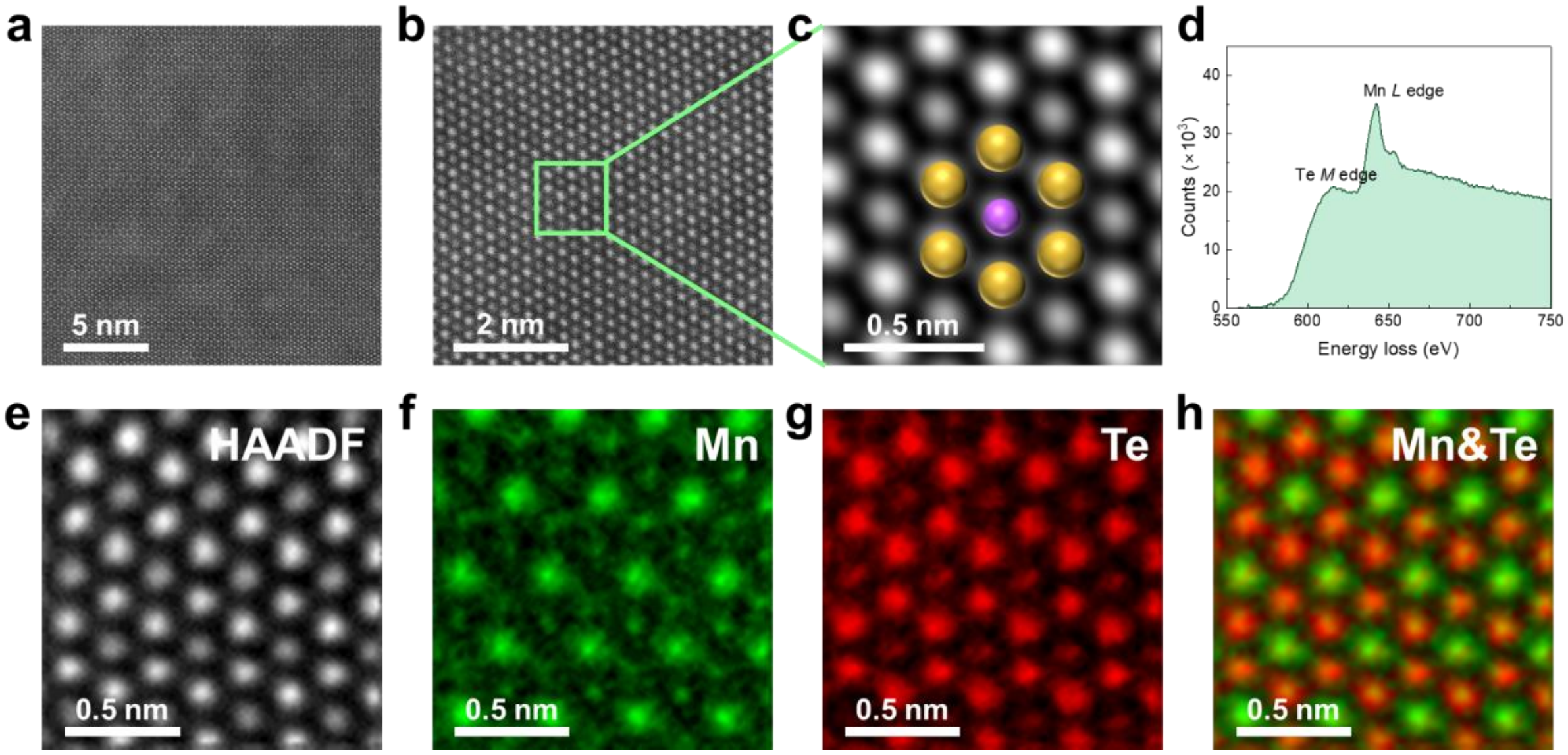

**Figure 2.** Atomic-resolution imaging and elemental analysis of MnTe single crystals. (a,b) STEM-HAADF images of MnTe captured along the [0001] zone axis, scale bars: (a) 5 nm, (b) 2 nm. (c) Magnified view of the boxed region in (b) overlaid with the atomic structure of MnTe (pink spheres: Mn; brown spheres: Te), scale bar: 0.5 nm. (d) EELS spectrum showing the Mn *L*-edge and Te *M*-edge (energy range: 550–750 eV). (e–h) Elemental mapping of MnTe: (e) HAADF image, (f) Mn map, (g) Te map, (h) Composite of Mn (green) and Te (red), Scale bars: 0.5 nm.

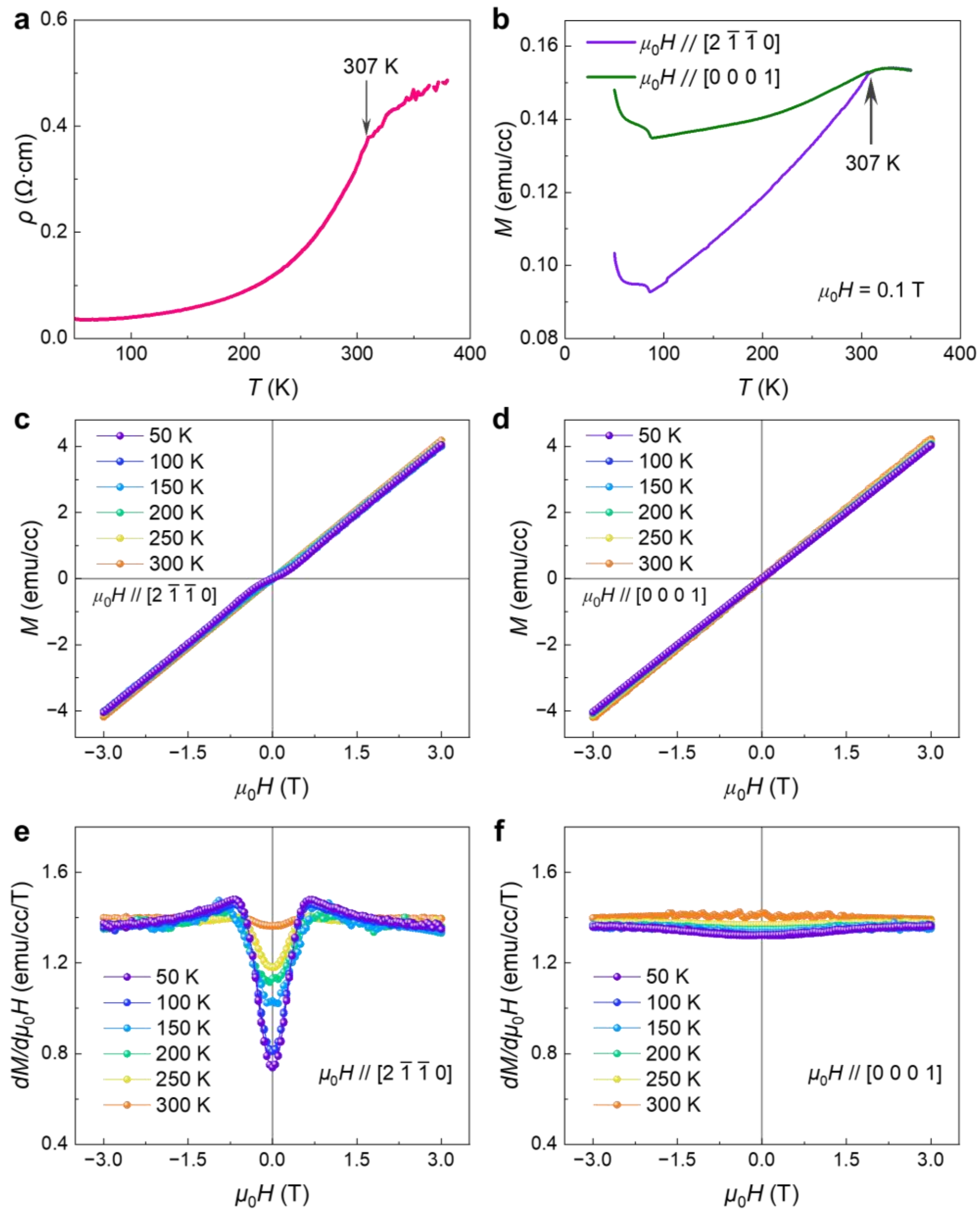

**Figure 3.** Electronic Transport and Magnetic Properties of MnTe. (a) Temperature-dependent resistance ($R$) of the MnTe sample from 50 to 380 K, showing an altermagnetic transition at 307 K. (b) Temperature-dependent magnetic moment ($M$) of the same sample measured from different crystallographic direction ([$2\bar{1}\bar{1}0$] and [0001]) from 50 to 380 K, with a corresponding transition at 307 K. (c) Magnetic-field dependent magnetization along the [$2\bar{1}\bar{1}0$] direction at various temperatures. (d) Magnetic-field dependent magnetization along the [0001] direction at various temperatures. (e) Derivative of magnetic-field dependent magnetization along the [$2\bar{1}\bar{1}0$] direction. (f) Derivative of magnetic-field dependent magnetization along the [0001] direction.

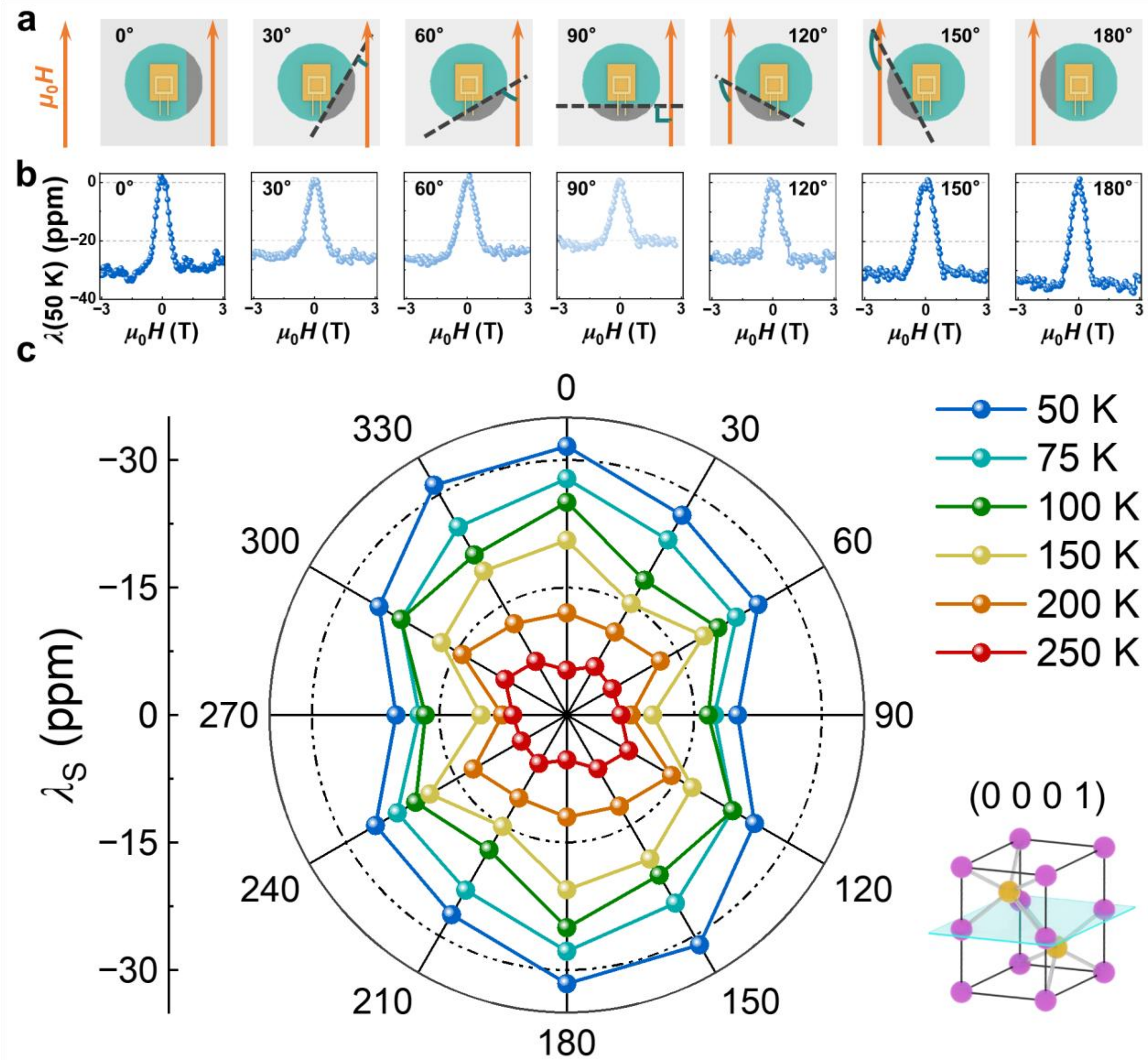

**Figure 4.** Anisotropic Magnetostriction in MnTe. (a) Illustration of the magnetostriction measurement configuration at selected angles from 0°, 30°, 60°, 90°, 120°, 150°, and 180°. The orange arrow indicates the direction of external magnetic field, $\mu_0 H$, and the turquoise plane represents the upper surface of MnTe sample. (b) Corresponding magnetostriction curves measured under ±3 T at 50 K of Sample #1. (c) Pole figure showing the angular dependence of the saturated magnetostriction for MnTe Sample #1, measured from 0° to 360° at temperatures of 50, 75, 100, 150, 200, and 250 K. The schematic in the bottom-right corner indicates the crystal orientation (0001) from which the pole figure was measured.

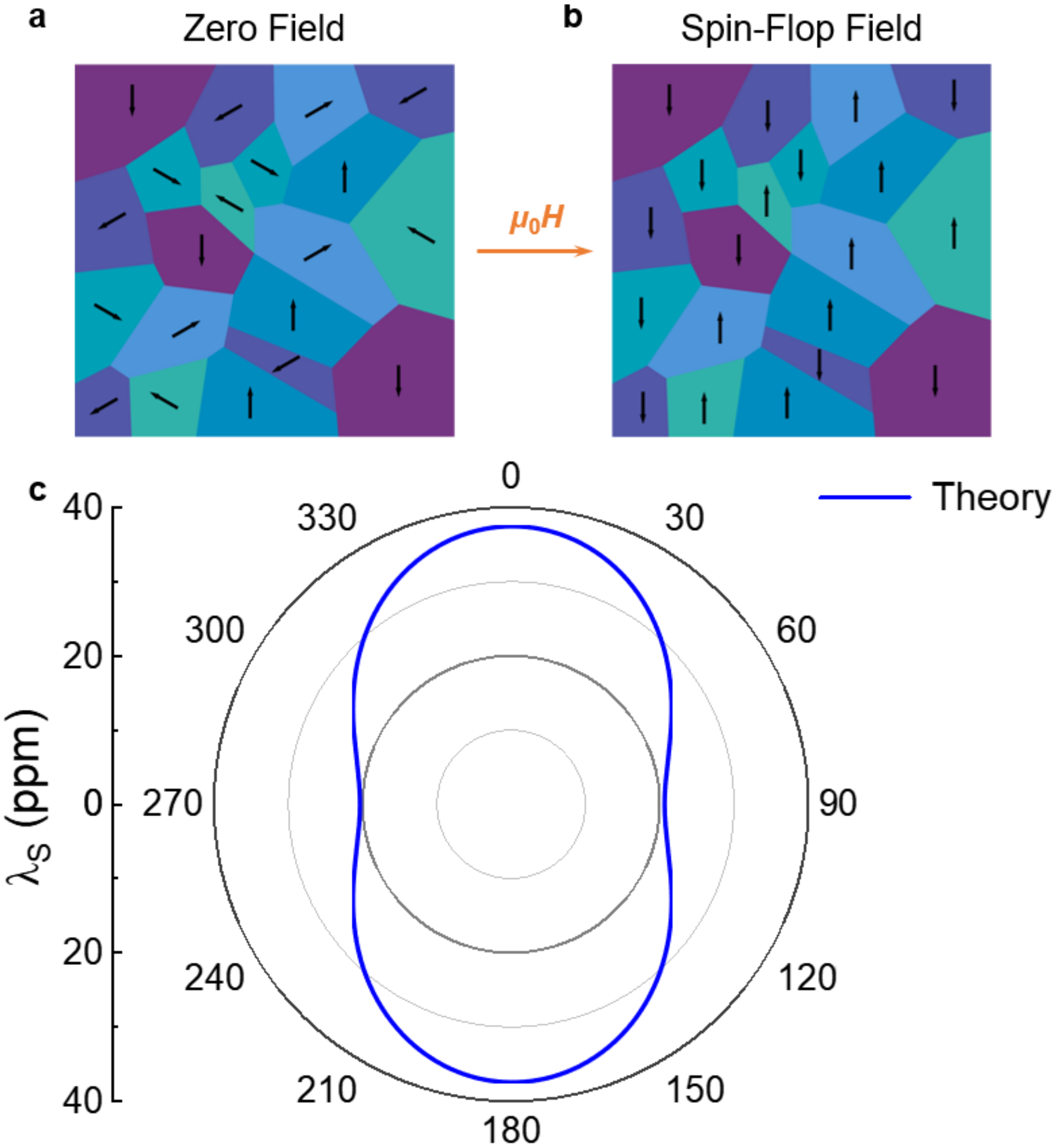

**Figure 5.** Multi-domain Picture of MnTe and Simulated Magnetostriction Anisotropy. (a) Schematic representation for the multi-domain picture of MnTe at zero magnetic field. Colored regions denote magnetic domains with distinct magnetic easy-axis orientations, and black arrows indicate the direction of the Néel vectors. (b) Corresponding schematic with an external field applied along $[2\bar{1}\bar{1}0]$ direction once the system reaches the spin-flop transition. The Néel vectors rotate in all domains, and at the spin-flop field the canted Néel vectors almost align the direction perpendicular to the magnetic field. (c) Simulated pole figure showing saturated magnetostriction as a function of angle $\theta$. For this case, the assumed domain fractions are in the ratio 4:2:2:5:2:2.

**TOC Graphic**

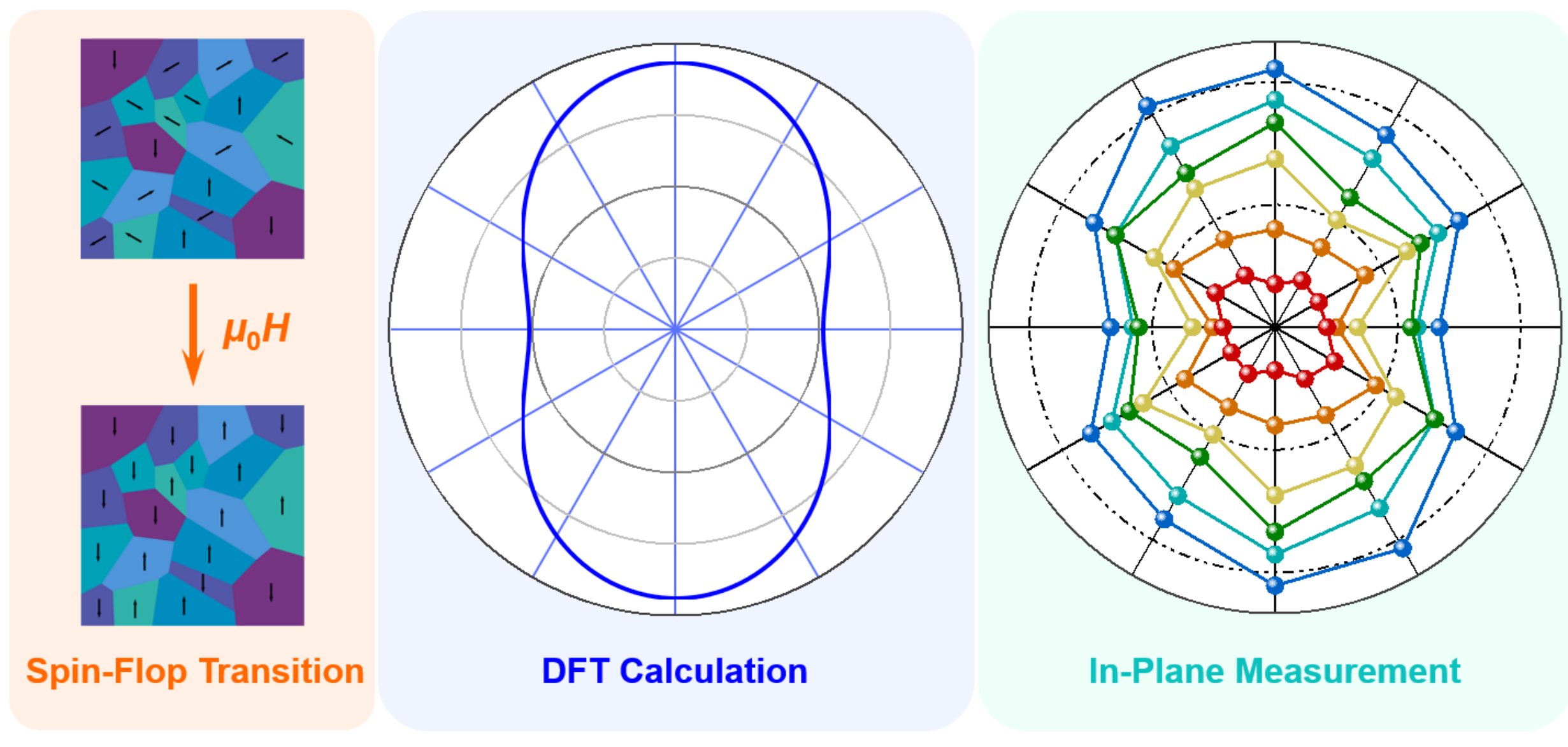

Saturated Magnetostriction in Altermagnet MnTe
μ0H
Spin-Flop Transition
DFT Calculation
In-Plane Measurement

*Supporting Information*

**Saturated and Anisotropic Magnetostriction in an Altermagnet**

*Zhiyuan Duan[1,8#], Qiyun Xu [2#], Peixin Qin[1,8]*, Li Liu[1,8], Guojian Zhao[1,8], Yuzhou He[3], Xiaoyang Tan[1,8], Sixu Jiang[1,8], Jingyu Li[1,8], Xiaoning Wang[4], Qinghua Zhang[3], Wenhui Duan[2,6,7], Yong Xu[2,6], Ziang Meng[1,8]*, Peizhe Tang[1,5]*, Chengbao Jiang[1,8]*, and Zhiqi Liu[1,8]**

[1]School of Materials Science and Engineering, Beihang University, Beijing, 100191, China.

[2]State Key Laboratory of Low Dimensional Quantum Physics and Department of Physics, Tsinghua University, Beijing, 100084, China.

[3]Beijing National Laboratory for Condensed Matter Physics, Institute of Physics, Chinese Academy of Sciences, Beijing, 100190, China.

[4]The Analysis & Testing Center, Beihang University, Beijing, 100191, China.

[5]Max Planck Institute for the Structure and Dynamics of Matter, Center for Free Electron Laser Science, Hamburg, 22761, Germany.

[6]Frontier Science Center for Quantum Information, Tsinghua University, Beijing, 100084, China.

[7]Institute for Advanced Study, Tsinghua University, Beijing, 100084, China.

[8]State Key Laboratory of Tropic Ocean Engineering Materials and Materials Evaluation, Beihang University, Beijing, 100191, China.

Corresponding author *Email: qinpeixin@buaa.edu.cn; mengza@buaa.edu.cn;

peizhet@buaa.edu.cn; jiangcb@buaa.edu.cn; zhiqi@buaa.edu.cn

**Experimental Section**

**Figures S1–S16**

**Table S1**

## Experimental Section

*Material preparation*

MnTe single crystals were grown using a simplified technique based on the Bridgman method. The synthesis process began with the preparation of quartz tubes suitable for high-temperature treatment. A mixture of 0.73 g of Manganese (99.95% purity) and 1.77 g of Tellurium (99.99% purity) powders was thoroughly mixed and ground in an agate mortar. The molar ratio of Mn and Te was set at 0.49:0.51 to reduce the melting point of flux. To avoid the reaction between MnTe and the quartz tube at high temperature, a custom-made conical boron nitride crucible was used to contain the mixture. The quartz tubes containing crucibles and mixture were sealed under a hydrogen flame with a base pressure of approximately $10^{-3}$ Pa. An OTF-1200X-S-VT vertical tubular furnace from HF-Kejing was employed for single crystal growth. Multiple quartz tubes were arranged vertically in the furnace to generate a stable, uniaxial temperature gradient necessary for crystal growth. The thermal schedule involved heating the furnace to 1150 °C for 15 hours, holding at this temperature for 10 hours, and then cooling gradually at a rate of 1 K/h down to 700 °C. The temperature was then held for another 10 hours, before a final cooling step to 300 °C at 5 K/h and subsequent natural cooling to room temperature. After sintering and cooling down, the obtained crystal was carefully smashed to find the cleaved crystal plane of (0001). Suitable crystals were selected for further cutting and polishing procedure.

*Structure characterization*

X-ray diffraction (XRD) patterns were obtained via a Bruker D8 Advance diffractometer with Cu-$K_\alpha$ radiation ($\lambda$ = 0.15418 nm). Transmission electron microscopy image was taken on a JEOL NeoArm system with an operating voltage of 200 kV. Electron backscatter diffraction (EBSD) data were collected using an Oxford Symmetry S3 detector. The measurements were carried out at a sample tilt of 70° relative to the horizontal, a working distance of 25 mm, an accelerating voltage of 25 kV, and a step size of 0.6 μm. Single-crystal XRD measurements were performed on a RIGAKU XtaLAB Synergy with Mo-$K_\alpha$ radiation ($\lambda$ = 0.71073 nm). Magnetic force microscopy (MFM) image was obtained on Bruker Dimension Icon.

*Magnetic and electrical transport*

Electrical transport measurements were performed using the resistance measurement option of a Quantum Design VersaLab system. Magnetic properties were characterized on the same instrument using the vibrating sample magnetometer (VSM) option.

*Magnetostriction Measurements*

Magnetostriction measurements were carried out using the rotator option within the Quantum Design VersaLab system. The detailed procedure is as follows: First, the as-grown MnTe single crystal was mechanically ground with sandpaper (up to 2000 grit) to create two parallel flat surfaces for strain gauge attachment. Second, the sample was then mounted onto the holder using GE varnish, a cryogenic-compatible adhesive. A strain gauge (Kyowa Electronic Instruments Co., Japan) was then attached to the top surface using a dedicated cyanoacrylate adhesive (CC-33A) from the same manufacturer. The bonding procedure was as follows: a drop of adhesive was placed on a plastic film. The strain gauge was then picked up with tweezers and gently dabbed onto the droplet to pick up an appropriate amount. Finally, it was placed onto the sample surface and pressed firmly with a finger for approximately 10 seconds until the adhesive was fully cured.

A critical aspect of this step is ensuring proper alignment: (i) the orientation of the strain gauge grid must be aligned with the intended magnetic field direction, and (ii) the crystallographic direction of interest within the sample must also be aligned with the field direction. Both alignments are essential for obtaining meaningful angular-dependent data.

In addition, the amount of adhesive should be carefully controlled to ensure full coverage of the metallic grid pattern for uniform strain transfer. In our tests, a slightly excessive amount of adhesive did not significantly affect the measured magnetostriction values, and repeated measurements yielded consistent results. However, insufficient adhesive coverage should be avoided as it may lead to inhomogeneous strain transfer.

The strain is detected via the change in electrical resistance of the gauge, which is measured using a four-probe method to minimize the effects of contact and lead resistances. The operating principle of a strain gauge is as follows. When the underlying sample surface stretches or compresses, the metallic foil within the gauge deforms accordingly. When the sample stretches, the wire becomes longer and narrower, leading to an increase in resistance. Conversely, when it compresses, the resistance decreases. To amplify this effect, the metallic foil is patterned into a serpentine (folded) geometry. This design increases the effective length of the wire along the strain axis, thereby enhancing the resistance change for a given strain. The relationship between the measured resistance change and the strain is given by:

$$\varepsilon = \frac{\Delta R}{K \cdot R_0}$$

Where $\Delta R$ stands for the change of resistance, $R_0$ stands for the original resistance of strain gauge. $K$ is a factor determined by the strain gauge, which is provided by the manufacturer. In general, the value of $K$ is 2.

During the magnetostriction measurement procedure, the crystallographic $[2\bar{1}\bar{1}0]$ direction was defined as 0°. For each measurement angle, a fresh strain gauge attachment was performed to ensure that the strain gauge orientation was precisely aligned with the intended direction relative to the $[2\overline{1}\overline{1}0]$ axis. The measurement was then carried out with the external magnetic field applied along the same direction, ensuring that the strain gauge always measured the magnetostriction parallel to the field ($\lambda_{//}$). After completing the measurement at a given angle, the sample and holder were immersed in ethanol to dissolve the GE varnish and detach the sample from the holder. The sample, with the strain gauge still attached, was subsequently transferred to acetone to dissolve the adhesive, allowing the strain gauge to be peeled off. The sample was then cleaned and prepared for the next angle, repeating the entire mounting and bonding procedure described above. This process was repeated for each angle from 0° to 180° in 30° increments within the (0001) plane. A schematic of the measurement configuration is provided in Figure S6.

*Group Theory Analysis*

The generic lowest-order magnetoelastic coupling in antiferromagnets can be written as $\epsilon_{ij}n_k n_l$, as discussed in previous studies.[1,2] For the $D_{6h}$ point group, the in-plane strain transforms either as $A_{1g}$, amounting to volume changes $\epsilon_{xx} + \epsilon_{yy}$, and $E_{2g}$, with doublet $(\epsilon_{xx} - \epsilon_{yy}, 2\epsilon_{xy})$, while the dipole magnetic order components $(n_x, n_y)$ transform as $E_{1g}$. So, for in-plane magnetoelastic coupling, the point-group symmetry allows two independent invariants: $(\epsilon_{xx} + \epsilon_{yy})(n_x^2 + n_y^2)$ and $(\epsilon_{xx} - \epsilon_{yy})(n_x^2 - n_y^2) + 4\epsilon_{xy}n_x n_y$.

*Theoretical Calculations*

DFT calculations were performed via Vienna Ab-initio Simulation Package,[3] using the exchange correlational functional of Perdew-Burke-Ernzerhof parametrized generalized gradient approximation.[4] A plane-wave energy cutoff of 400 eV and a Monkhorst-Pack k-mesh of 21× 21× 10 were adopted. All calculations were performed considering the noncollinear magnetism, while the spin-orbit coupling was included. DFT+U method was employed with U= 4 eV and J= 0.97 eV parameters taken from a similar compound of manganese.[5] The relaxed lattice constants are $a = b = 4.201$ Å and $c = 6.696$ Å. A series of ±0.3%, ±0.6%, and ±1.0% $\epsilon_B$ or $\epsilon_\mu$ strains were applied, yielding the magnetic ground-state energies along the $[2\bar{1}\bar{1}0]$ and $[01\bar{1}0]$ directions as listed in Table S1; the elastic coefficients and coupling strength were subsequently obtained by fitting, yielding $c_{11}$= 81.6 GPa, $c_{12}$= 28.2 GPa, $\lambda_1 N_0^2 = 2.42\times10^{-2}$ GPa, $\lambda_2 N_0^2 = 1.54\times10^{-3}$GPa.

## REFERENCES

(1) Simensen, H. T.; Troncoso, R. E.; Kamra, A.; Brataas, A. Magnon-Polarons in Cubic Collinear Antiferromagnets. *Phys. Rev. B* **2019**, *99* (6), 064421.

(2) Steward, C. R. W.; Fernandes, R. M.; Schmalian, J. Dynamic Paramagnon-Polarons in Altermagnets. *Phys. Rev. B* **2023**, *108* (14), 144418.

(3) Baral, R.; Abeykoon, A. M. M.; Campbell, B. J.; Frandsen, B. A. Giant Spontaneous Magnetostriction in MnTe Driven by a Novel Magnetostructural Coupling Mechanism. *Adv. Funct. Mater.* **2023**, *33* (46), 2305247.

(4) Perdew, J. P.; Burke, K.; Ernzerhof, M. Generalized Gradient Approximation Made Simple. *Phys. Rev. Lett.* **1996**, *77* (18), 3865–3868.

(5) Antropov, V. P.; Antonov, V. N.; Bekenov, L. V.; Kutepov, A.; Kotliar, G. Magnetic Anisotropic Effects and Electronic Correlations in MnBi Ferromagnet. *Phys. Rev. B* **2014**, *90* (5), 054404.

**Figure S1**

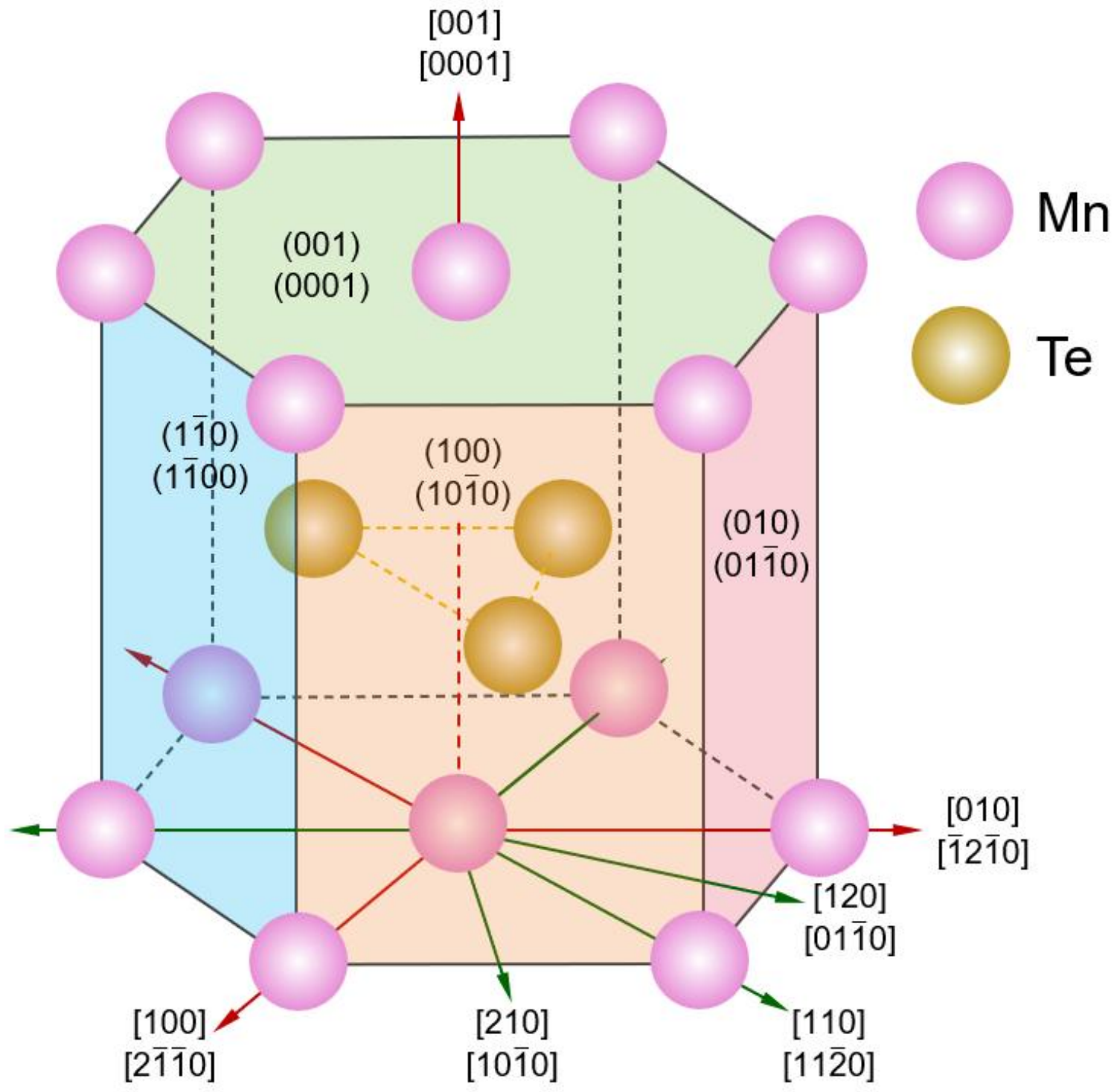

**Figure S1.** Atomic illustration of the MnTe hexagonal lattice, highlighting the relationship between three-index Miller notation [*uvw*] and four-index Miller-Bravais notation [*uvtw*] for representative directions.

**Figure S2**

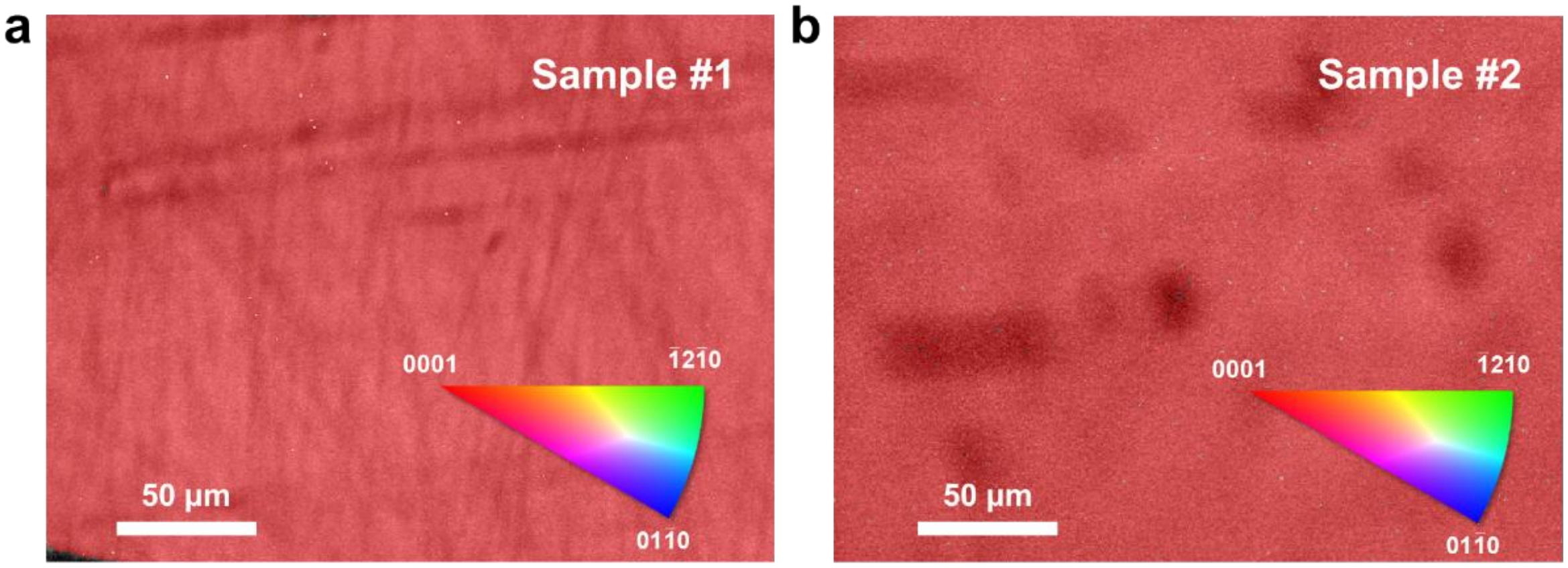

**Figure S2**. EBSD IPF map of the MnTe crystal (Z-direction). The red color corresponds to (0001), with no detectable grain boundaries. (a) Sample #1, scale bar: 50 μm, (b) Sample #2, scale bar: 50 μm.

**Figure S3**

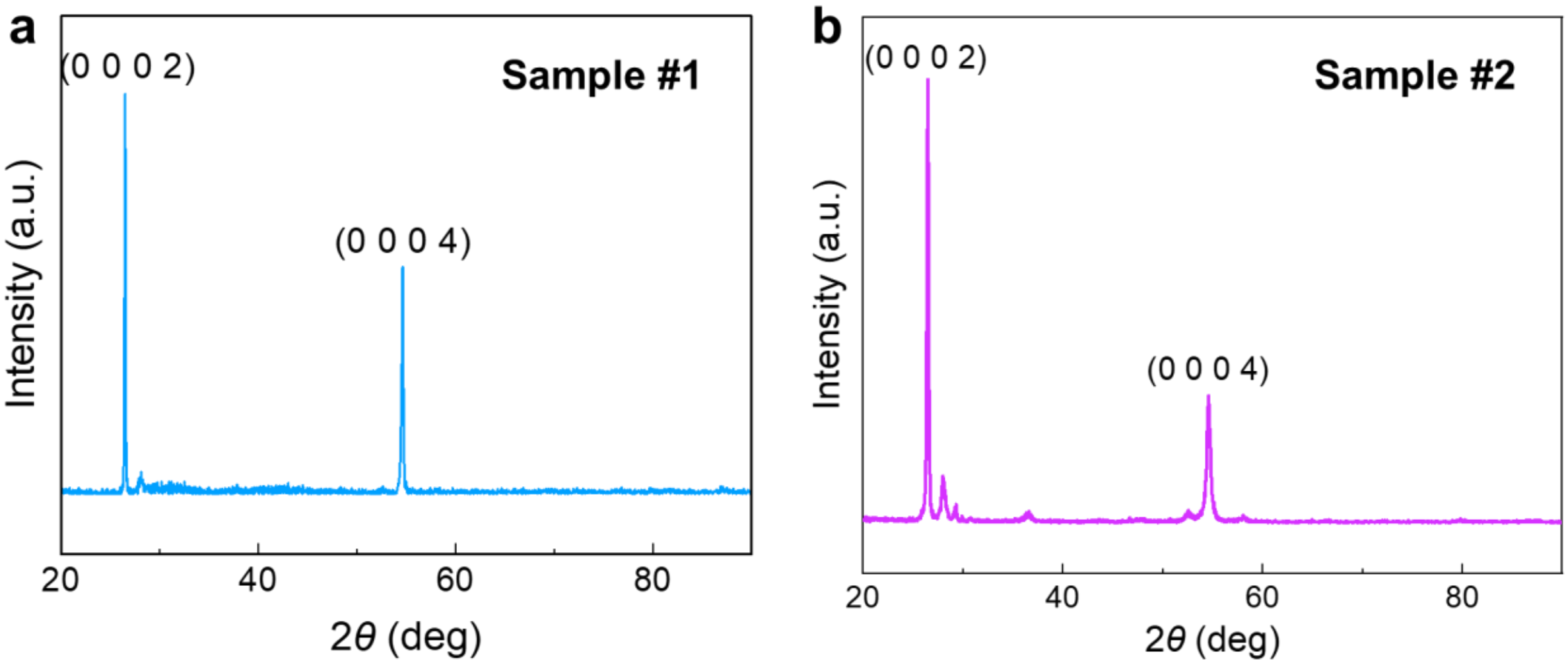

**Figure S3**. X-ray diffraction (XRD) pattern of the measured surface of the MnTe crystal. (a) Sample #1, (b) Sample #2.

**Figure S4**

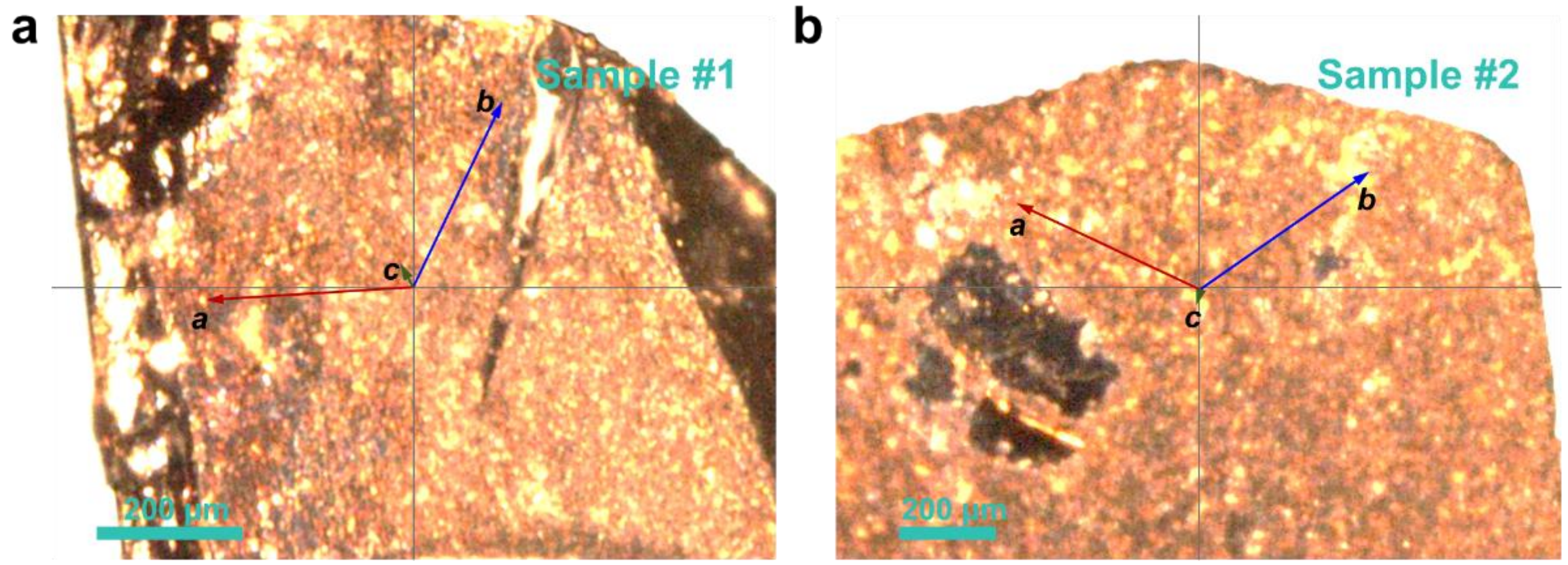

**Figure S4**. Single-crystal X-ray diffraction (SC-XRD) of the measured surface of the MnTe single crystal. (a) Sample #1, scale bar: 200 μm, (b) Sample #2, scale bar: 200 μm.

**Figure S5**

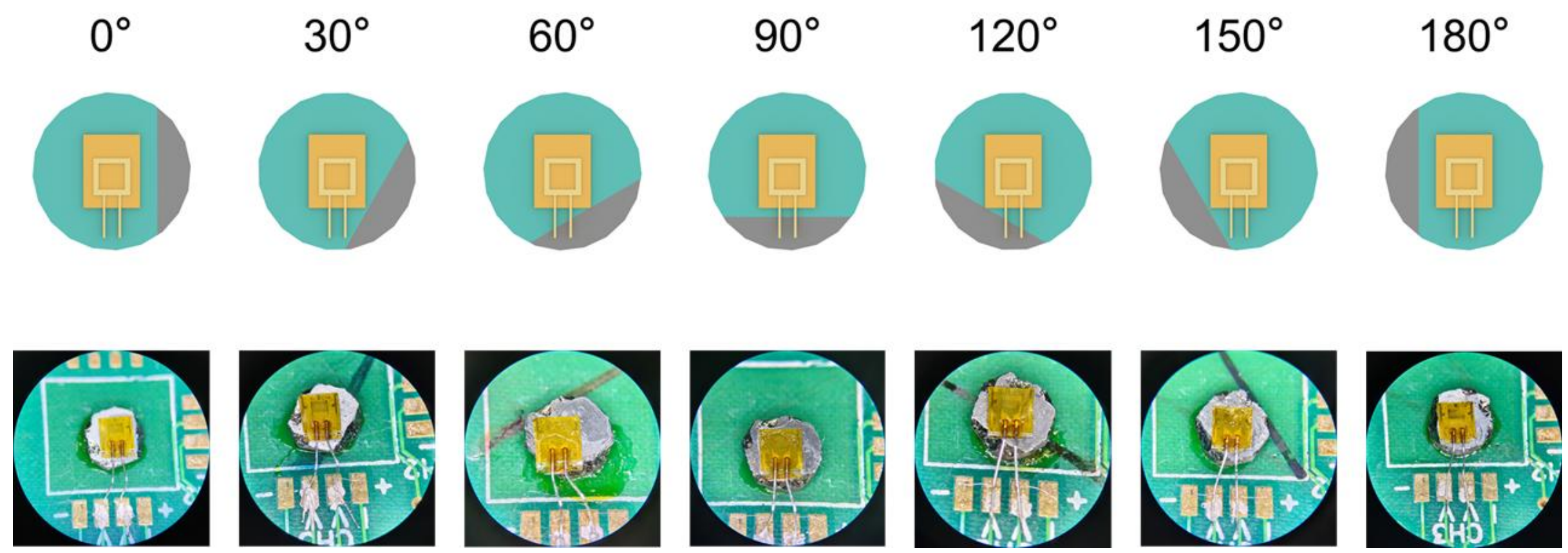

**Figure S5**. Illustration and corresponding optical image of the magnetostriction measurement (Sample #1). The resistance of the strain gauge was measured via a four-probe method.

**Figure S6**

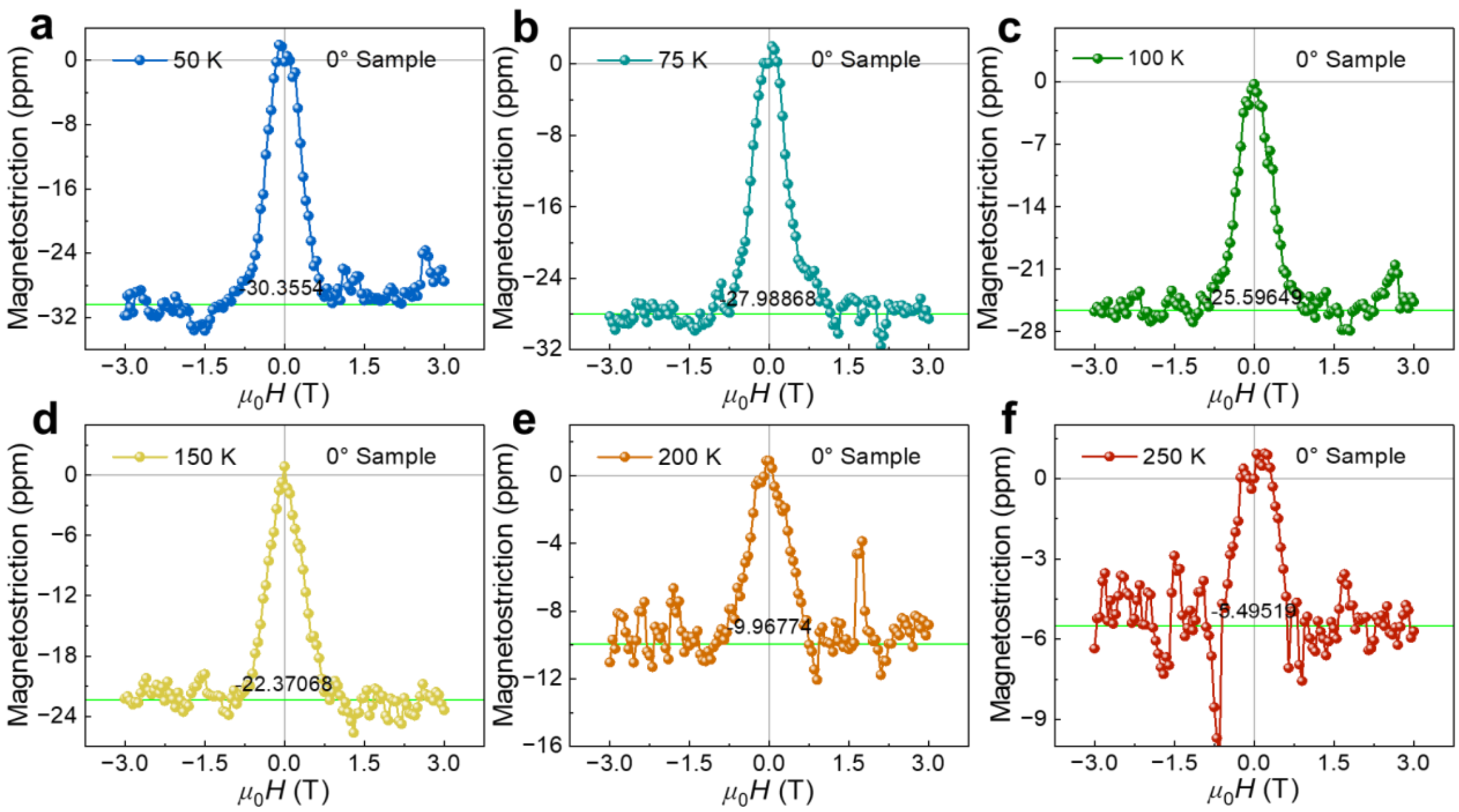

**Figure S6.** Magnetic field dependence of magnetostriction for Sample #1 in the (0001) crystal plane along the 0° ([$2\bar{1}\bar{1}0$]) direction, measured at (a) 50, (b) 75, (c) 100, (d) 150, (e) 200, and (f) 250 K, respectively.

**Figure S7**

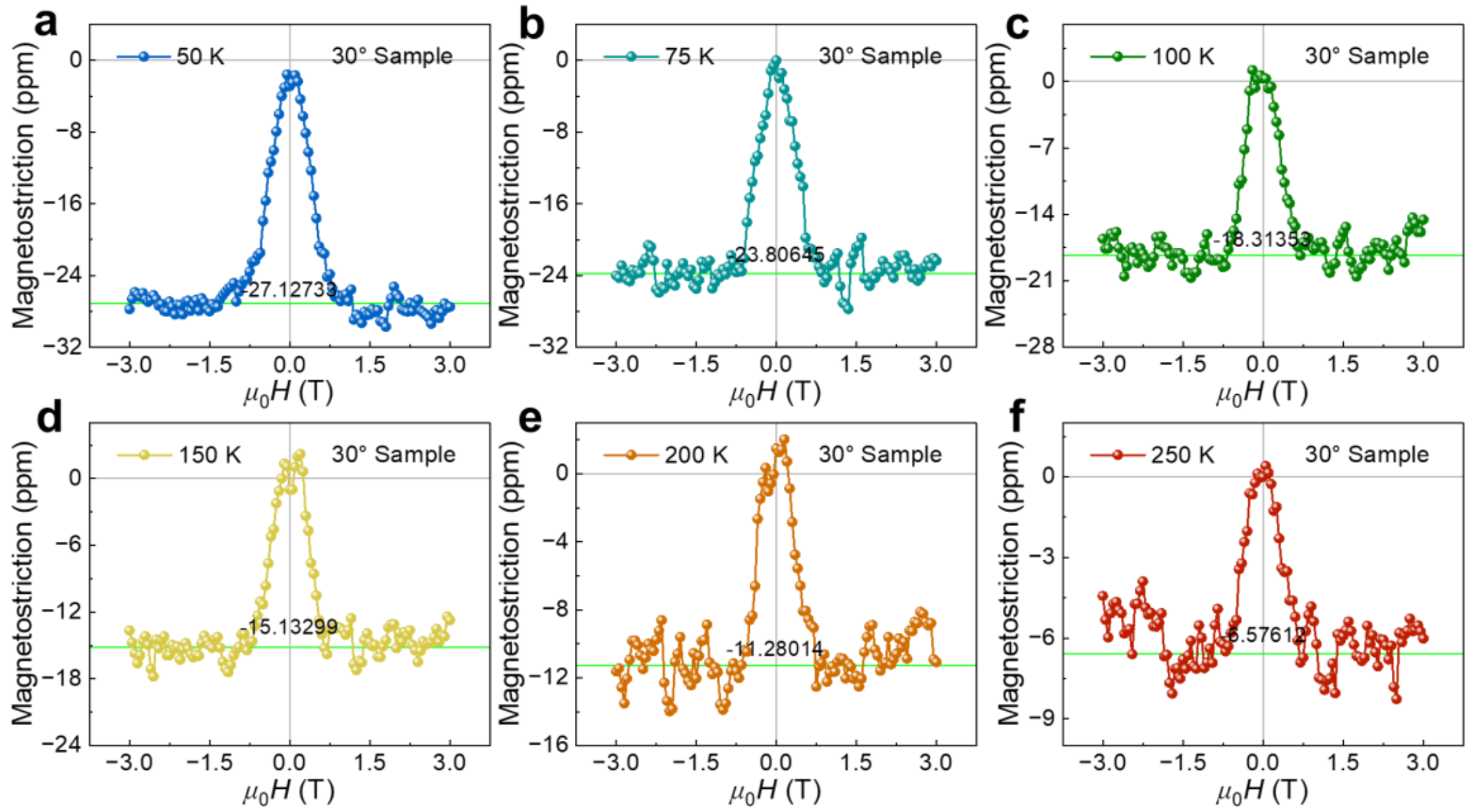

**Figure S7.** Magnetic field dependence of magnetostriction for Sample #1 in the (0001) crystal plane along the 30°direction, measured at (a) 50, (b) 75, (c) 100, (d) 150, (e) 200, and (f) 250 K, respectively.

**Figure S8**

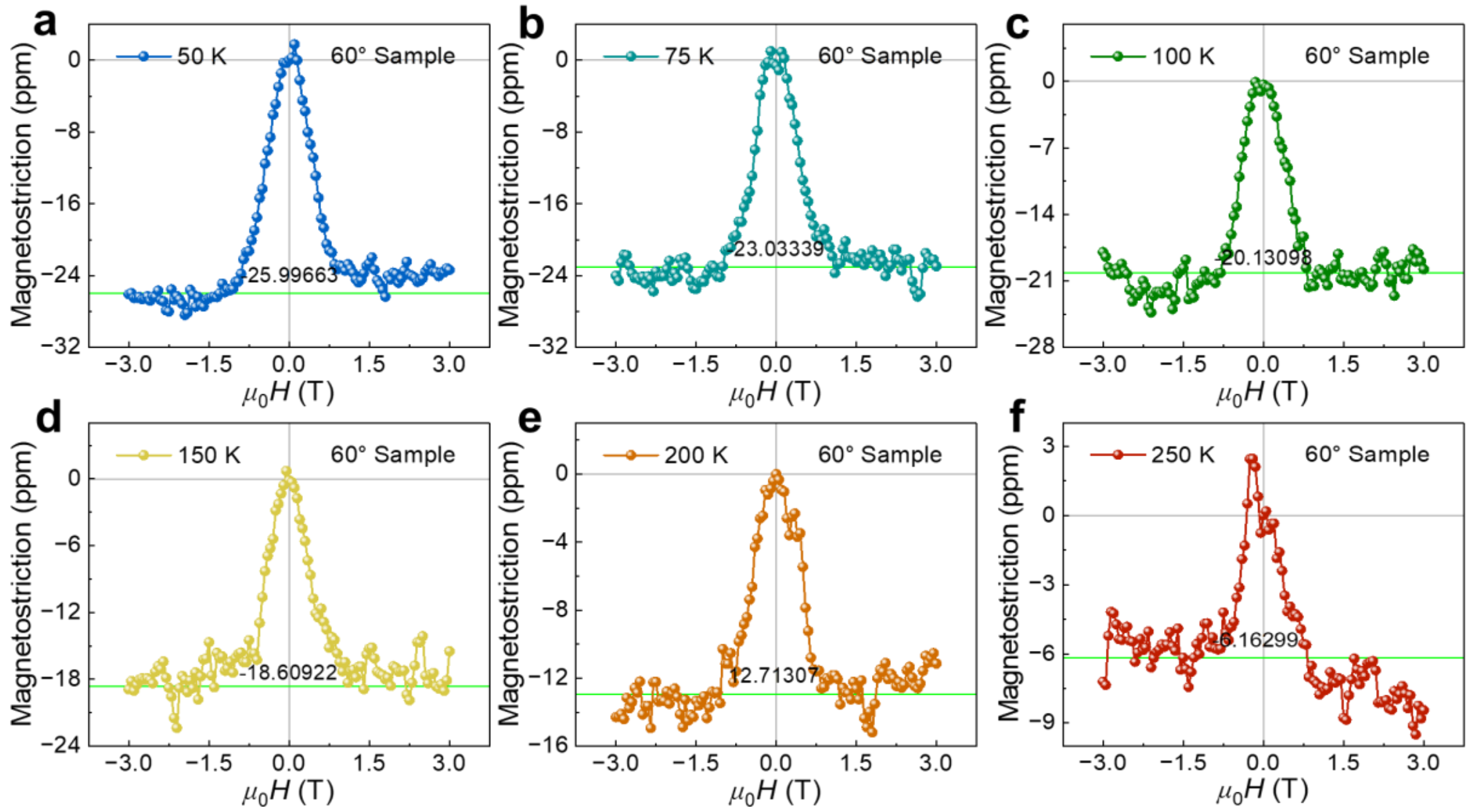

**Figure S8.** Magnetic field dependence of magnetostriction for Sample #1 in the (0001) crystal plane along the 60°direction, measured at (a) 50, (b) 75, (c) 100, (d) 150, (e) 200, and (f) 250 K, respectively.

**Figure S9**

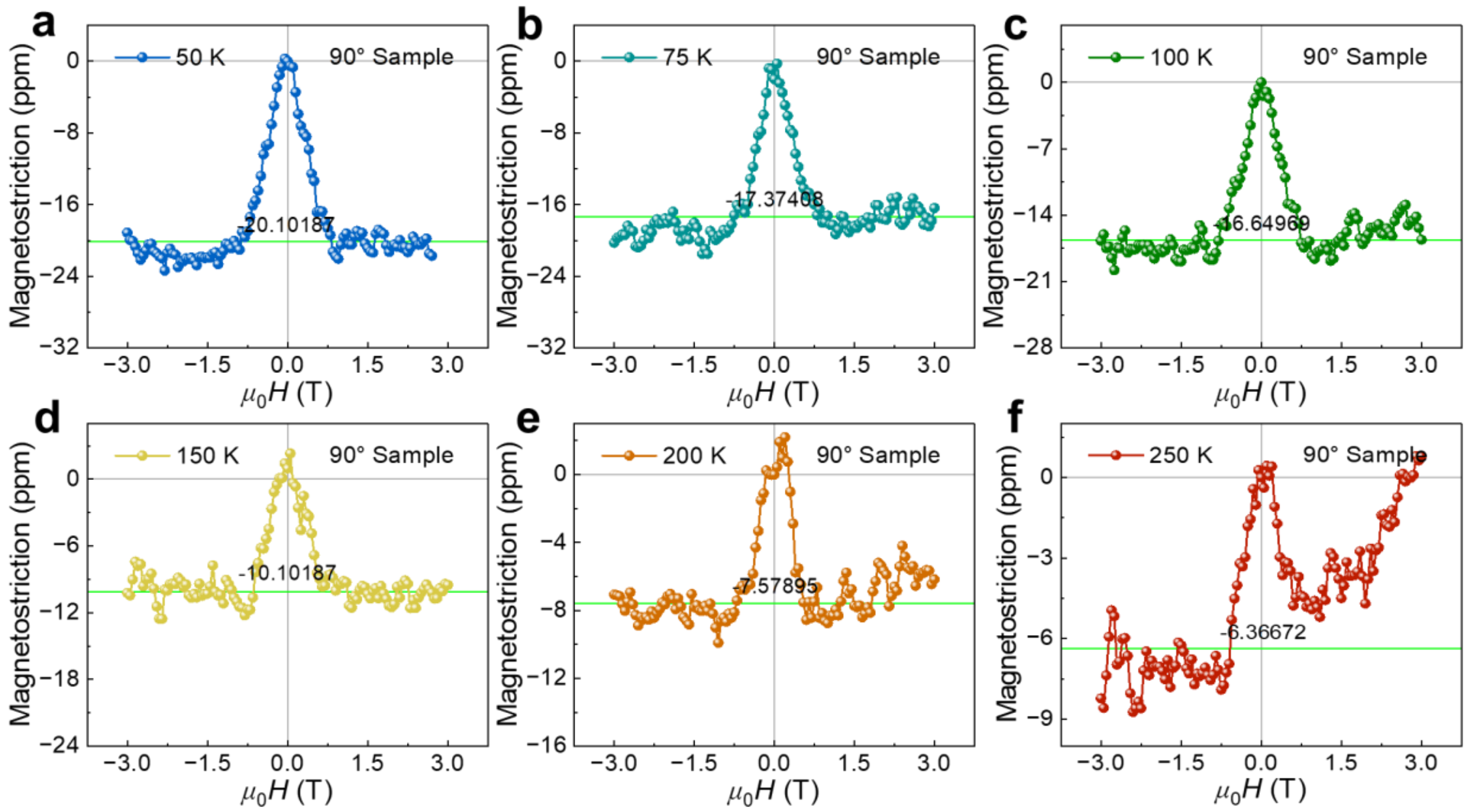

**Figure S9.** Magnetic field dependence of magnetostriction for Sample #1 in the (0001) crystal plane along the 90° ([01$\bar{1}$0]) direction, measured at (a) 50, (b) 75, (c) 100, (d) 150, (e) 200, and (f) 250 K, respectively.

**Figure S10**

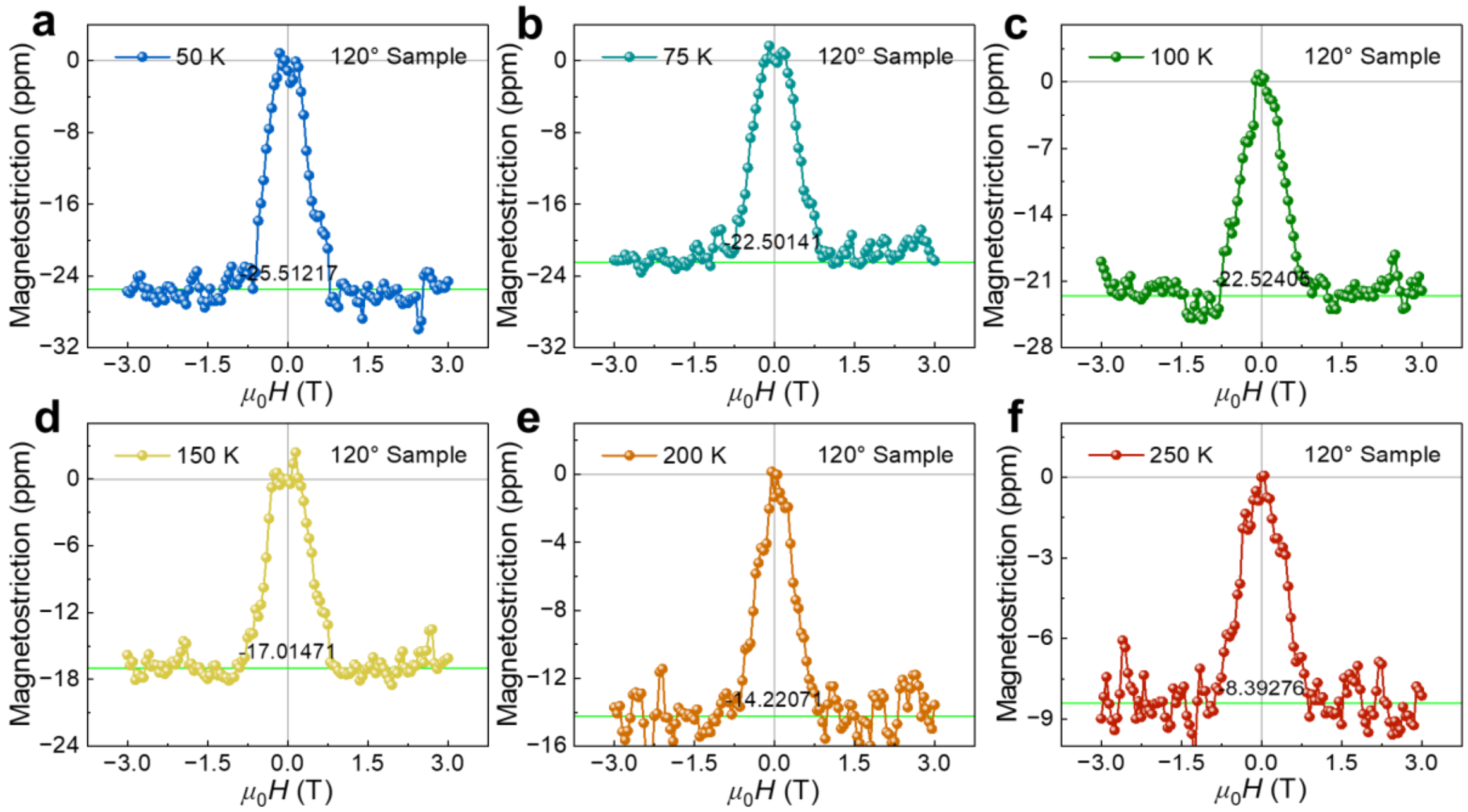

**Figure S10.** Magnetic field dependence of magnetostriction for Sample #1 in the (0001) crystal plane along the 120°direction, measured at (a) 50, (b) 75, (c) 100, (d) 150, (e) 200, and (f) 250 K, respectively.

**Figure S11**

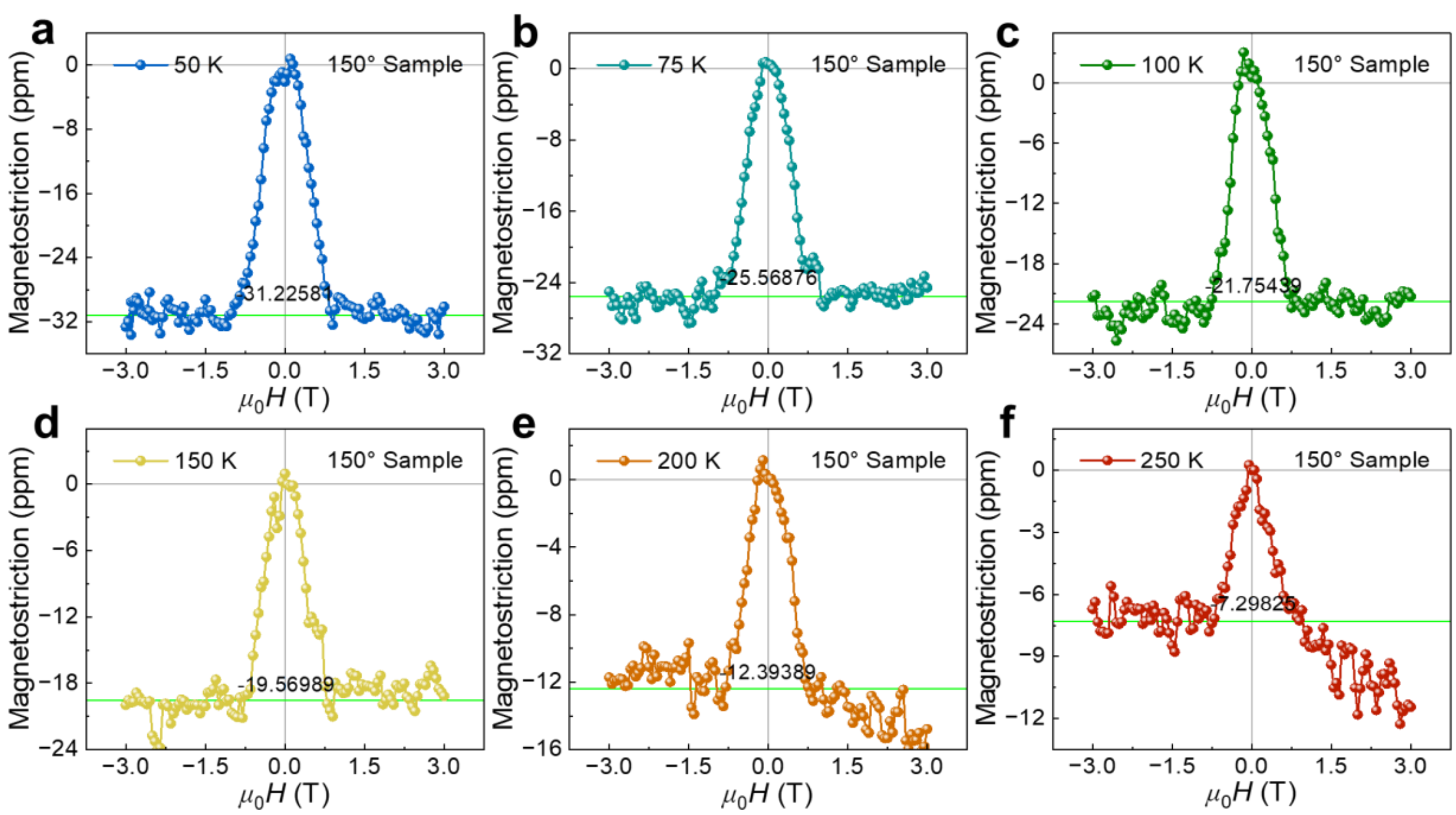

**Figure S11.** Magnetic field dependence of magnetostriction for Sample #1 in the (0001) crystal plane along the 150°direction, measured at (a) 50, (b) 75, (c) 100, (d) 150, (e) 200, and (f) 250 K, respectively.

**Figure S12**

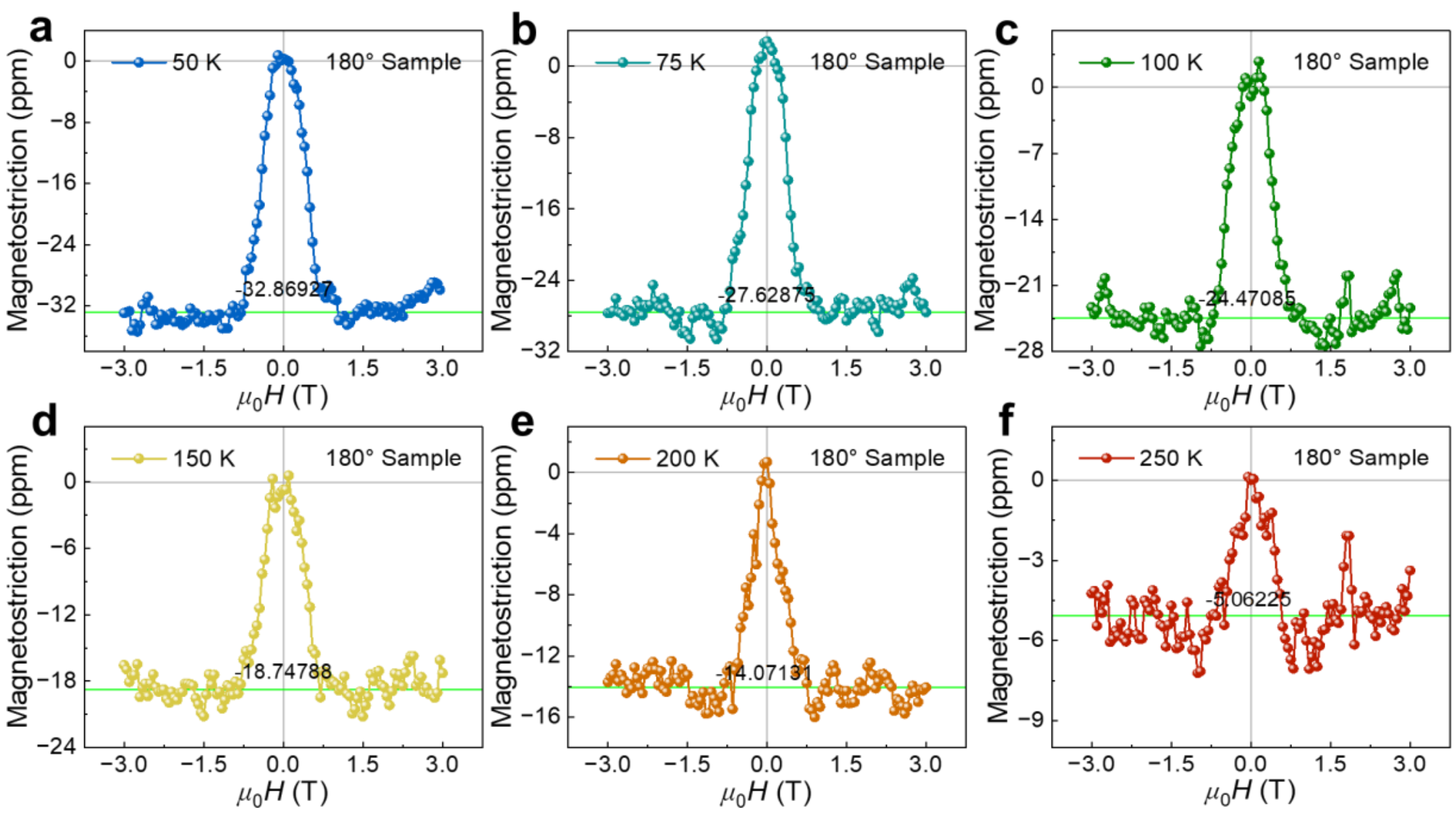

**Figure S12.** Magnetic field dependence of magnetostriction for Sample #1 in the (0001) crystal plane along the 180°direction, measured at (a) 50, (b) 75, (c) 100, (d) 150, (e) 200, and (f) 250 K, respectively.

**Figure S13**

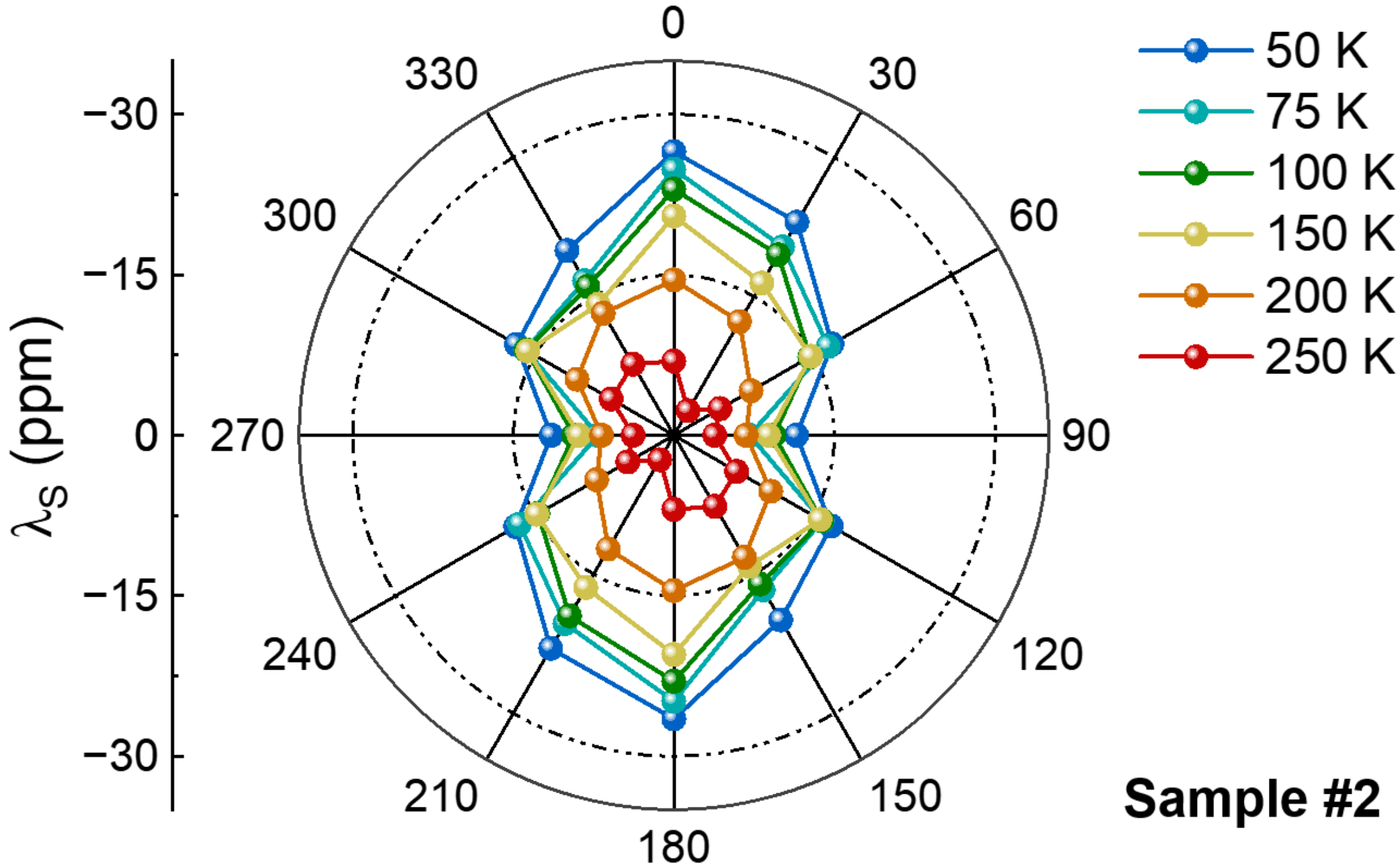

**Figure S13.** Pole figure of the saturated magnetostriction of Sample #2, showing the angular dependence. measured from 0° to 360° at temperatures of 50, 75, 100, 150, 200, and 250 K.

**Figure S14**

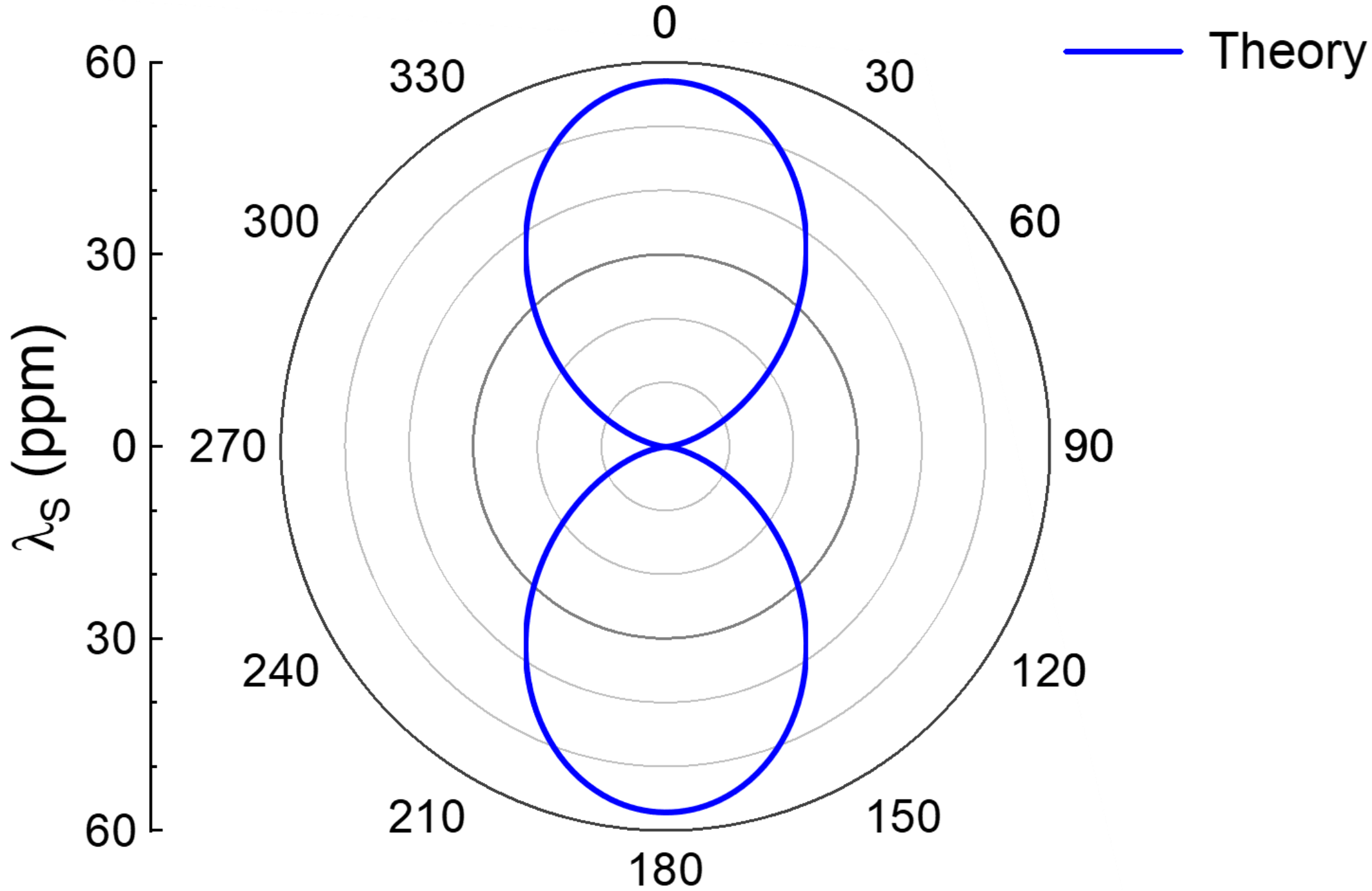

**Figure S14.** Pole figure illustrating the theoretical angular dependence of the saturated magnetostriction in single-domain MnTe. The angle denotes the difference between the field direction and the Néel vector.

**Figure S15**

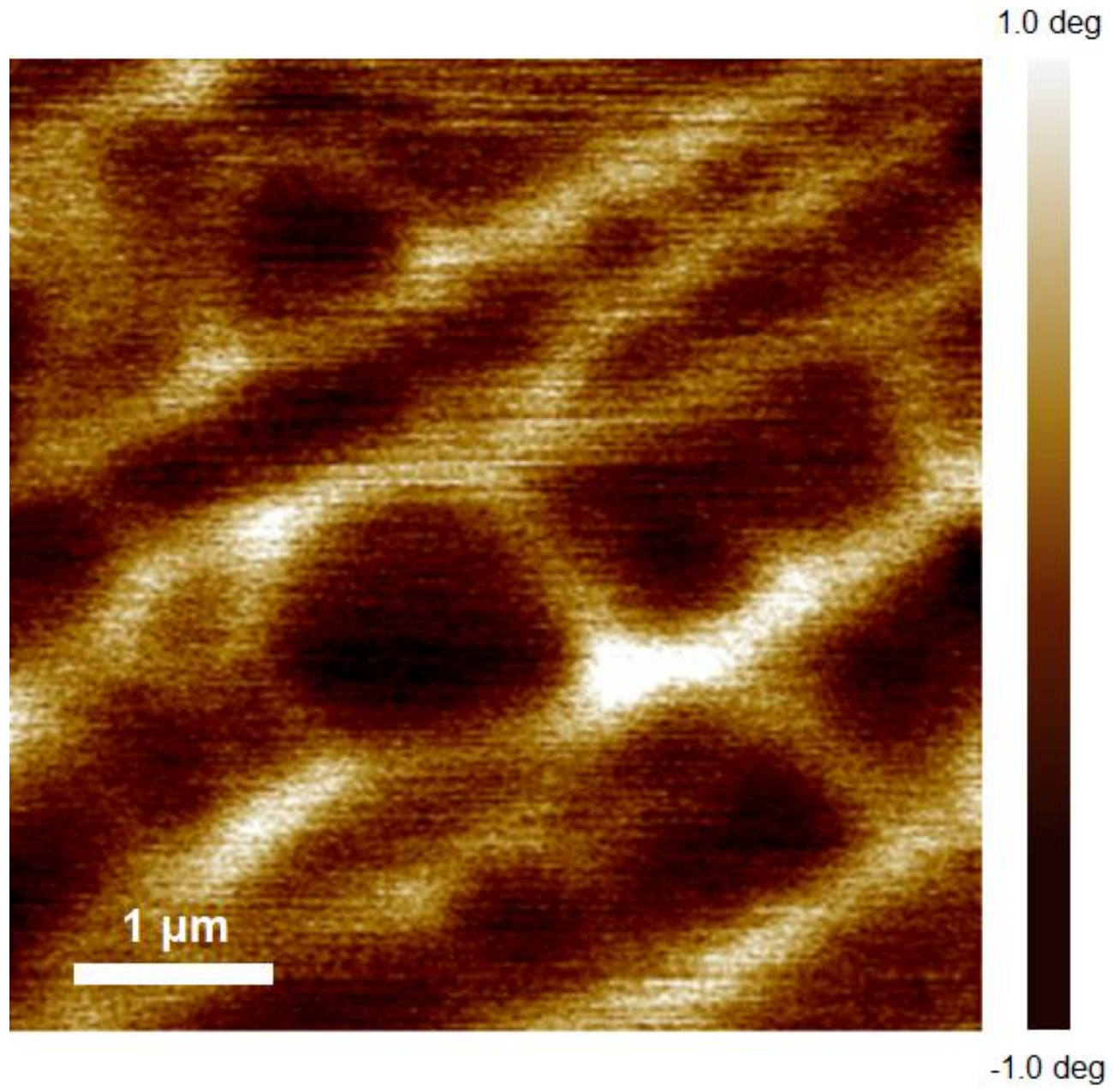

**Figure S15.** MFM topographic image of the MnTe measurement surface (scan area: 5 µm × 5µm), scale bar: 1µm.

**Figure S16**

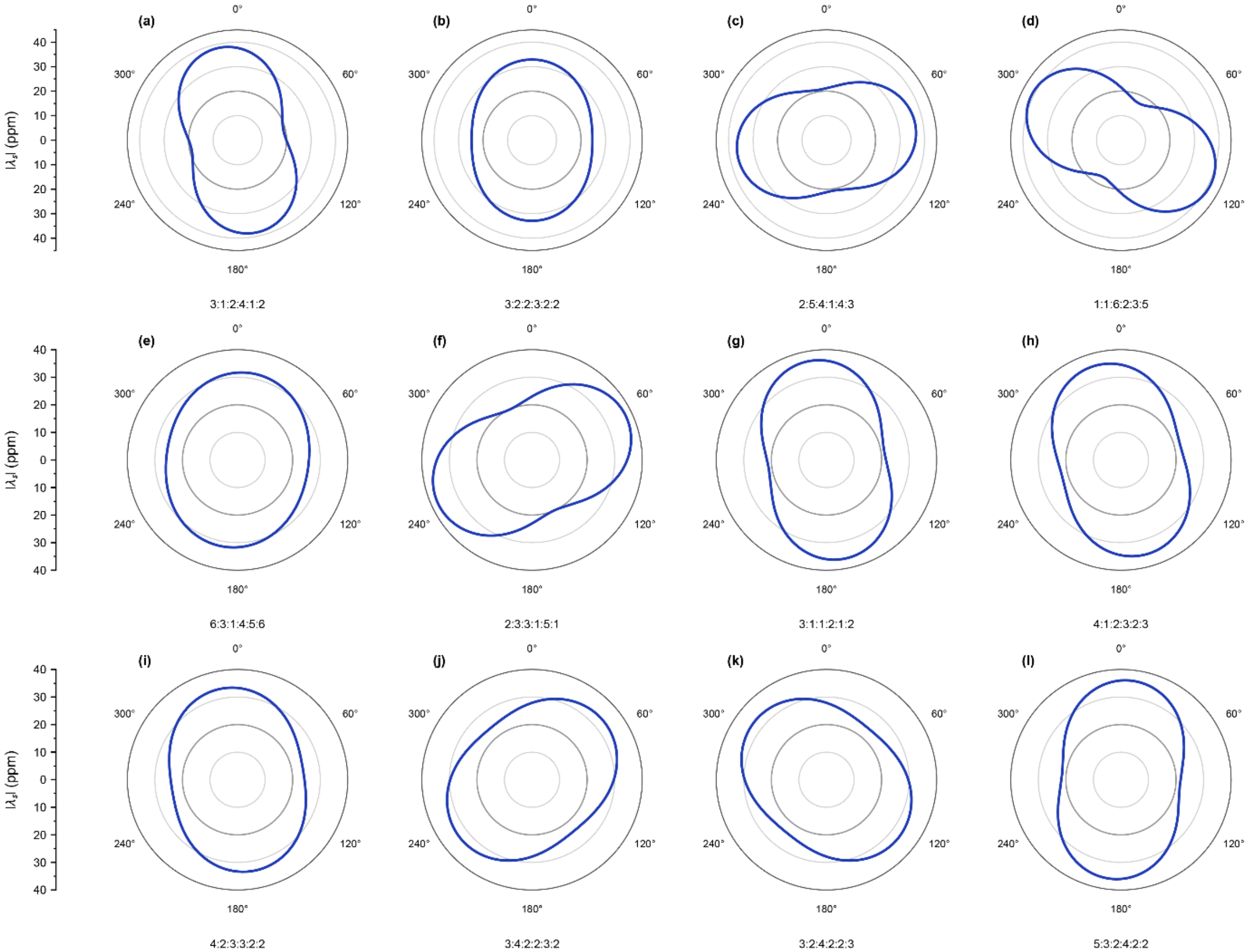

**Figure S16**. Pole figures of the saturated magnetostriction for representative multidomain population sets in MnTe. Panels (a–l) correspond to different domain fractions (ratios shown below each panel). The six components in each ratio are assigned to the six in-plane symmetry-equivalent domain orientations generated by C6 rotational symmetry. The simulations show that the two-fold anisotropic pattern is preserved over broad variations of domain fractions, while the signal amplitude depends on the specific ratio.

**Table S1**

**Table S1**. DFT calculations for the total free energy in meV per unit cell for different lattice strain. $\Delta E(\epsilon) = E(\epsilon) - E(\epsilon = 0)$ represents the relative energy difference after applying strain, and the Néel vector direction is indicated in the subscript of the energies.

| $\epsilon_B/\%$ | -1.0 | -0.6 | -0.3 | 0.3 | 0.6 | 1.0 |
|---|---|---|---|---|---|---|
| $\Delta E_{[100]}(\epsilon_B)/\text{meV}$ | 1.928 | 0.689 | 0.207 | 0.114 | 0.547 | 1.606 |
| $\Delta E_{[120]}(\epsilon_B)/\text{meV}$ | 1.891 | 0.665 | 0.188 | 0.062 | 0.529 | 1.550 |

| $\epsilon_\mu/\%$ | -1.0 | -0.6 | -0.3 | 0.3 | 0.6 | 1.0 |
|---|---|---|---|---|---|---|
| $\Delta E_{[100]}(\epsilon_\mu)/\text{meV}$ | 0.896 | 0.327 | 0.086 | 0.122 | 0.315 | 0.873 |
| $\Delta E_{[120]}(\epsilon_\mu)/\text{meV}$ | 0.818 | 0.253 | 0.032 | 0.076 | 0.283 | 0.824 |